\documentclass[fleqn,usenatbib]{mnras}

\usepackage[T1]{fontenc}

\DeclareRobustCommand{\VAN}[3]{#2}
\let\VANthebibliography\thebibliography
\def\thebibliography{\DeclareRobustCommand{\VAN}[3]{##3}\VANthebibliography}

\usepackage{graphicx}	
\usepackage{amsmath}	
\usepackage[linesnumbered,boxed]{algorithm2e}
\usepackage{xcolor}
\SetKwRepeat{Do}{do}{while}

\usepackage{amssymb}	
\usepackage{mathtools}

\usepackage{newtxtext,newtxmath}

\DeclarePairedDelimiter{\ceil}{\lceil}{\rceil}
\newcommand{\V}[1]{\mathcal{V}_{pq, t\nu}}
\newcommand{\VAP}[1]{V_{pq, t_m\nu_m}}
\newcommand{\VA}[1]{\mathcal{V}_{pq, t_m\nu_m}}

\title[Big data and compression]{Lossy Compression of Large-Scale Radio Interferometric Data}

\author[Atemkeng et al.]{
M Atemkeng,$^{1}$\thanks{E-mail: m.atemkeng@gmail.com (MA)}
S Perkins,$^{2}$
E Seck,$^{3}$
S Makhathini,$^{2,4,5}$
O Smirnov,$^{2,4}$
L Bester,$^{2,4}$
B Hugo,$^{2,4}$
\\
$^{1}$Department of Mathematics, Rhodes University, Grahamstown 6139, South Africa\\
$^{2}$South African Radio Astronomy Oberservatory, Black River Park, 2 Fir Street, Observatory, Cape Town, 7925, South Africa\\
$^{3}$African Institute for Mathematical Sciences (AIMS), Limbe, Cameroon\\
$^{4}$ Department of Physics and Electronics, Rhodes University, PO Box 94, Grahamstown, 6140, South Africa\\
$^{5}$School of Physics, University of the Witwatersrand, 1 Jan Smuts Avenue Johannesburg, 2001, South Africa
}

\date{Accepted XXX. Received YYY; in original form ZZZ}

\pubyear{2015}

\begin{document}
\label{firstpage}
\pagerange{\pageref{firstpage}--\pageref{lastpage}}
\maketitle

\begin{abstract}
Radio telescopes produce vast amounts of data and the data volume is set to increase as upcoming telescopes (e.g. SKA, ngVLA) come online. The vast amounts of data are an important issue to deal with in the context of calibration, deep wide-field imaging, storage and archiving for sub-sequence processing. This work proposes to reduce visibility data volume using  a baseline-dependent lossy compression technique that preserves smearing at the edges of the field-of-view. We exploit the relation of  the rank of a matrix and the fact that  a low-rank approximation can describe the raw visibility data as a sum of basic components where each basic component corresponds to a specific Fourier component of the sky distribution. As such, the entire visibility data is represented as a collection of data matrices from baselines, instead of a single tensor. This has the benefit of parallel computation and allows the investigation of baseline-dependent rank$-r$ decomposition. The proposed methods are formulated as follows: provided a large dataset of the entire visibility data; the first algorithm, named \textit{simple SVD} projects the data into a regular sampling space of rank$-r$ data matrices.  In this space, the data for all the baselines has the same rank, which makes the compression factor equal across all  baselines. The second algorithm, named \textit{BDSVD} projects the data into an irregular sampling space of rank$-r_{pq}$ data matrices. The  subscript $pq$ indicates that the rank of the data matrix varies across baselines $pq$, which makes the compression factor baseline-dependent. MeerKAT and the European Very Long Baseline Interferometry Network are used as reference telescopes  to evaluate and compare the performance of the proposed methods against traditional methods, such as traditional averaging and baseline-dependent averaging (BDA).  For the same spatial resolution threshold, both \textit{simple SVD} and \textit{BDSVD} show effective compression by two-orders of magnitude higher than   traditional averaging and BDA. At the same space-saving rate, there is no decrease in spatial resolution and  there is a reduction in the noise variance in the data which improves the S/N to over $1.5$ dB at the edges of the field-of-view. The proposed compression methods offer superior compression but requires processing data at full data resolution, for reduced implementation complexity.
\end{abstract}




\section{Introduction}
\label{Sect:introduction}
\newcommand{\Vtm}{\boldsymbol{V}}
\newcommand{\VV}{\mathcal{V}}
\newcommand{\NN}{\mathcal{N}}
\newcommand{\VVV}{V}
\newcommand{\WW}{W}
\newcommand{\I}{\boldsymbol{I}}
\newcommand{\etapq}{\boldsymbol{\eta}}
\newcommand{\FF}{\mathcal{F}}
\newcommand{\CC}{\boldsymbol{C}}
\newcommand{\AAAA}{\boldsymbol{A}}
\newcommand{\SIGMA}{\boldsymbol{\Sigma}}
\newcommand{\aaa}{\boldsymbol{a}}
\newcommand{\cc}{\boldsymbol{c}}
\newcommand{\LAMBDA}{\boldsymbol{\Lambda}}
\newcommand{\vvv}{\boldsymbol{v}}
\newcommand{\OO}{\boldsymbol{0}}
\newcommand{\uu}{\boldsymbol{u}}
\newcommand{\II}{\mathcal{I}}
\newcommand{\llll}{\boldsymbol{l}}
\newcommand{\SIGMACAl}{\mathcal{\Sigma}}

Radio interferometric arrays consist of an assembly of radio antennas that are correlated in pairs to produce complex data values, known as visibilities or sampled spatial frequencies. The data volume grows quadratically with the number of antennas, very large sky surveys, high spectral and temporal resolutions. Processing and storing this volume of visibility data has become a challenge, for example calibrating  and translating the data to the image space via resampling and fast Fourier transform operations. The large volume of visibility data is an important problem to deal with in the context of calibration and deep wide-field imaging with current radio interferometric arrays such as MeerKAT~\citep{jonas2009meerkat}, ASKAP~\citep{johnston2008science}, NenuFAR~\citep{zarka2015nenufar}, LOFAR~\citep{van2013lofar} and future radio interferometers, including the Square Kilometre Array (SKA)\citep{dewdney2009square}.  To resolve the morphological structure of compact  sources,  the SKA will sample very high spatial frequencies with long baselines.   This high spatial frequency sampling will take the SKA to an unprecedented visibility data volume era; this will require more computation and improved data compression strategies. Long-term data archiving of calibrated products is necessary, and will benefit from the implementation of new data compression algorithms.

Data compression is critical to reduce the costs involved in storing, processing and archiving data. Many data compression strategies exist, however, choosing a relevant strategy depends on the science case and how the information related to the science case is encoded in the data. Therefore, a high data compression rate requires that we understand the information in the data and the science case. For a given dataset it is possible that only a few data points contain the majority of information relevant to the science, while other points contain noise or information irrelevant to the case \citep{bobin2008compressed}.

In visibility data, for example, the noise from system electronics, signal from unwanted areas of the sky and spatial frequencies from redundant baselines can be removed from the data. There are mathematical tools that can break down this type of dataset into a new form that makes it easier to understand how the information in the dataset is represented based on some degree of importance. The properties of each of these mathematical tools for breaking down the dataset are different and their suitability depends on the science case.  Additional factors must also be considered when choosing a compression strategy, to find the best tradeoff between the compression ratio, signal loss and science.

In practice, the visibility data are integrated and averaged over finite time and frequency intervals which introduces decorrelation effects. To limit decorrelation, the finite time and frequency intervals are kept very small so that the phase term on long baselines is preserved. The decorrelation attenuates and changes the morphology of sources at the edges of the field-of-view (FoV). If the time and frequency intervals are scaled up to some limit, for a given FoV, then traditional averaging can  be used to reduce the data volume. 
Deconvolution routines must correctly take into account the smearing effects to avoid limiting the dynamic range of the image; this is discussed in details by several authors~\citep{cotton2009effects,atemkeng2016using, tasse2018faceting, atemkeng2020fast}. 

Since high-resolution sampling of the spatial frequency is  only required on long baselines, baseline-dependent averaging (BDA) methods have recently gained popularity and can be used to compress the visibility data and suppress sources out of the FoV while maintaining spectral and temporal resolutions on the long baselines. While  BDA can potentially offer compression capabilities on short baselines and maintain the high spectral and temporal resolution required on long baselines, it still requires further investigation.  The Measurement Set v2.0 (MS)\citep{kemball2000measurementset} format does not support storing variable frequency bins as imposed by BDA. The Measurement Set v2.0 format can, in principle, store variable frequency bins required by BDA, thus this approach requires the creation of a spectral window for each decomposition of the frequency domain.
Current calibration algorithms expect the data to be regularly sampled. However, to make use of  BDA compression capabilities, calibration algorithms will need to adapt, especially for the different spectral and temporal resolutions along baselines. It is also important to choose appropriate solution intervals for calibration when choosing BDA parameters. Well-defined interpolation windows and weighting schemes are needed for imaging. This is not a problem for imaging algorithms as they place unstructured samples on a regular grid.

Visibility data is frequently archived for later use; traditional averaging and/or  BDA is applied before archiving. However, many criteria must be considered when choosing compression parameters, as high spectral and temporal resolution is imperative for certain science cases (e.g. VLBI science, spectral line studies) and for calibration routines that take into account the variation of the primary beam (PB) and the effect of the ionosphere, for example.  Traditional averaging and/or BDA is accompanied by a decrease in the spectral and temporal resolution at least on shorter baselines for BDA and must therefore be applied with care. There is, therefore, a need to investigate data compression techniques that can archive visibility data with little to no decrease in spectral and temporal resolution. 

In this work, we use low-rank approximation for visibility data compression.
 A low-rank approximation minimises a loss function that measures the best-approximating data matrix with a reduced-rank relative to the original data matrix i.e. a low-rank approximation represents a higher-rank data matrix by a reduced-rank data matrix with little to no loss of information. A  data matrix, $\Vtm$ of size $M\times N$ has a  rank$-r$ approximation given by:
\begin{equation}
    \Vtm\simeq \AAAA \CC,
\end{equation}
where the data matrices $\AAAA$ and $\CC$ are of size $M\times r$ and $r\times N$, respectively. The number of entries in the rank$-r$ approximation is $r(M+N)$, which can now be stored in the place of  $\Vtm$ to save memory and/or computation since $r(M+N)$ is considerably smaller than $MN$ if $r$ is relatively small and  $M, N$ are relatively large. Thus, $1\leq r < MN/(M+N)$. $\mathcal{O}(r(M+N))$ disk space is now required to save the rank$-r$ approximation of $\Vtm$ rather than $\mathcal{O}(MN)$. 

As of now, in the big data regime, many fields applying image processing project a high-rank dataset to a lower-rank by  \textit{dimensionality reduction}. That is a process of feature (dimension)  extraction that preserves the most relevant information (reduction) in the data. Dimensionality reduction primarily functions as a pre-processing step to reduce the need for high computing resources, and is classified into two different classes; linear and nonlinear.
The most well-known low-rank approximation that belongs to the linear class of dimensionality reduction algorithms is the singular value decomposition (SVD). The SVD decomposes the data into separate sets of relevant features, noisy and redundant components.  The properties of the SVD are fully explored in other fields. It is popular in signal processing due to its orthogonal matrix decomposition. Several authors use SVD to tackle different tasks with interferometric data; for example \citet{offringa2010post} discussed using SVD to separate strong radio frequency interference from weak astronomical signals. To speed up imaging and deconvolution, \citet{vijay2017fourier} used SVD to compress interpolated and gridded visibility data. Recent work proposes using holographic measurement of the PB and SVD  to deal with spatial and spectral variation of the PB; e.g. \citet{iheanetu2019primary} discussed the possibility of determining a sparse representation of the PB with a few features and \citet{ansah2020denoising}  discussed mitigating noise in the PB by choosing the strongest components of the decomposition. 

The interest in exploiting the SVD to compress the visibility data is not accidental; high-sensitivity radio interferometric arrays are dominated by short baselines: SVD exploits redundancies in short baselines and can isolate noise in the data. A rank$-r$ SVD decomposition of a data matrix of size $M\times N$ requires $ \mathcal{O}(rMN)$ operations: $ \mathcal{O}(N^3)$ in the case of a square data matrix \citep{gu1996efficient,demmel2007fast,shishkin2019fast}. Computation of this magnitude is impractical in a big data regime. However, the SVD methods described in this paper should not be confused with a method that treats the entire visibility data as a single tensor and where there is an assumption of similarity between different baselines. We exploit the relation of the SVD to the rank of a matrix and the fact that the SVD describes the raw visibility data as a sum of components, where each component corresponds to a specific Fourier component of the sky distribution.  As such, the entire visibility data is represented as a collection of matrices derived for each baseline, instead of a single tensor. This facilitates baseline-dependent rank$-r$ decomposition that can be performed independently for each baseline, allowing for parallel computing.   The proposed methods are formulated below. Given a dataset of visibility data:
\begin{itemize}
    \item    The first algorithm projects the data into a regular sampling space of rank$-r$ data matrices.  In this space, the data for all the baselines has the same rank, which makes the compression factor equal across all  baselines. We referred to this algorithm as \textit{simple SVD}. The \textit{simple SVD} is shown to be effective in compressing the visibility data by two-orders of magnitude higher than  traditional averaging with a reduction in the noise variance in the data.
    \item The second algorithm projects the data into an irregular sampling space of rank$-r_{pq}$ data matrices. The extra subscript $pq$ indicates that the rank of the data matrix varies across baselines, which makes the compression factor baseline-dependent.   
    We referred to this method as baseline-dependent SVD (\textit{BDSVD}).
\end{itemize}
It should be noted that \textit{simple SVD} and \textit{BDSVD} are used differently compared to traditional averaging and BDA. The compressed visibility data from traditional averaging and BDA are used directly as input to other tasks such as imaging or calibration, with a lower computation cost as opposed to the uncompressed data. This leads to a decrease in spectral and temporal resolution. The \textit{simple SVD} and \textit{BDSVD}  are used to archive the visibility data and the data must be decompressed before subsequent use. This has the advantage of maintaining the spectral and temporal resolution of the original data. The use cases are summarised in Table \ref{table:1}. In this paper, we implement the \textit{simple SVD} and \textit{BDSVD} serially and provide detailed justification for parallel implementation. Whereafter we suggest a parallel algorithm that can reduce the computation from $\mathcal{O}\big(N_bMN^2)$ to $\mathcal{O}\big(N_bMN^2/N_p)$; where  $N_b$ and $N_p$ are the number of baselines and the number of computing nodes, respectively. The implementation of this algorithm is not part of this work, we leave it for future work. 

The rest of the paper is organised as follows: Section~\ref{background} gives an overview of radio  interferometric measurements and the theory behind  SVD.  Section~\ref{methods} describes the mathematical formalism behind the algorithms. Section~\ref{Effectsnoise} analyses the effects of the proposed  algorithms on the image and the noise. Simulation results are discussed in Section~\ref{simulation} and Section~\ref{conclusion} draws conclusions and suggests future works. Our mathematical notations are summarised in Table \ref{table:x}.

\begin{table*}
\centering
\begin{tabular}{|p{3cm}| p{2cm} |p{9.cm} |p{3.2cm}|} 
 \hline
 \textbf{Methods} &  \textbf{Compression} & \textbf{Preserved spectral and temporal resolution within the FoV?}& \textbf{Speedup processing?} \\ [0.5ex] 
 \hline\hline
 \textbf{Traditional averaging} & yes & no & yes \\ 
 \textbf{BDA} & yes & no & yes \\
 \textbf{\textit{Simple SVD}} & yes & yes & no \\
 \textbf{\textit{BDSVD}} & yes & yes & no \\
 \hline
\end{tabular}
\caption{The use cases of traditional averaging and BDA vs. \textit{simple SVD} and \textit{BDSVD}}
\label{table:1}
\end{table*}

\section{Background}
\label{background}
In this section, we discuss the mathematical framework of the visibility data measurements provided by the radio interferometric measurement equation (RIME) formalism~\citep{hamaker1996understanding, smirnov2011revisiting}. The framework is used to understand, describe and measure the performance of  \textit{simple SVD} and \textit{BDSVD}.

\subsection{Full-sky RIME and traditional averaging}
\label{RIMEaveraging}
Under the RIME formalism and for a single point source, \citet{smirnov2011revisiting} establishes the relationship between a measured visibility at  time $t$ and frequency $\nu$ and the averaged visibility over some time and frequency integration interval $[t_0, t_1]\times [\nu_0, \nu_1]$. We extend this  relationship for a full-sky RIME formalism:
\begin{align}
\VV(\uu_{pq}(t, \nu)) &=\sum_{\llll} \big(\mathcal{G}_{p\llll t\nu}\II_{\llll}\mathcal{G}_{q\llll t\nu}^\dagger\big)\mathcal{K}^{\llll}_{pq t\nu}.\label{eq1}
\end{align}
The scalar term $\mathcal{K}^{\llll}_{pq t\nu}=\text{e}^{-2i\pi (\llll-\llll_0)\uu_{pqt\nu}}$
describes the effects of the position of antennas $p, q$ and the rotation of the baseline vector $\uu_{pq}(t, \nu)\equiv\uu_{pqt\nu}$ which tracks a source located in the direction of the unit vector $\llll$. A compensating delay $\llll_0$ is introduced by the correlator to enforce that  $\mathcal{K}^{\llll}_{pqt\nu}\equiv 1$ at the phase centre of the observation:
\begin{align*}
    &\uu_{pqt\nu}=\frac{\nu}{c}\begin{bmatrix}
    u_{pqt}  \\
    v_{pqt} \\
    w_{pqt}
\end{bmatrix}, 
\llll=\begin{bmatrix}
    l  \\
    m \\
    n
\end{bmatrix},
\llll_0=\begin{bmatrix}
    0  \\
    0 \\
    1
\end{bmatrix}.
\end{align*}
In the above equations,  $^\dagger$ is the complex transpose operator, $c$ stands for the speed of light, $\II_{\llll}$ is the sky brightness, $\mathcal{G}_{p\llll t\nu}$ and $\mathcal{G}_{q\llll t\nu}^\dagger$ group the direction-dependent Jones matrices of antenna $p$ and $q$, respectively.

Below is an approximate of the average of Eq.~\ref{eq1} over $[t_0, t_1]\times [\nu_0, \nu_1]$:
\begin{align}
\VVV_{pq (t\nu)_{i}} &\simeq\sum_{\llll} \text{sinc}\frac{\Delta \Phi}{2}\text{sinc}\frac{\Delta \psi}{2}\big( \mathcal{G}_{p\llll (t\nu)_{i}}\II_{\llll}\mathcal{G}_{q\llll (t\nu)_{i}}^\dagger\big)\mathcal{K}^{\llll}_{pq (t\nu)_{i}},\label{eq:ap}
\end{align}
where $(t\nu)_{i}=t_i\nu_i$ with $t_i=(t_1+t_0)/2$ and $\nu_i= (\nu_1+\nu_0)/2$. The phase differences,  $\Delta \Phi$ and $\Delta \psi$ are defined as:
\begin{align*}
    \Delta \Phi&= \text{arg}\mathcal{G}_{p\llll t_{1}\nu_{i}}  +\text{arg}\mathcal{G}_{q\llll t_{1}\nu_{i}}^\dagger+\text{arg}\mathcal{K}^{\llll}_{pq t_{1}\nu_{i}}\\
               &~~~~~~~~~~~~~~~~~~~~-\Big( \text{arg}\mathcal{G}_{p\llll t_{0}\nu_{i}}+\text{arg}\mathcal{G}_{q\llll t_{0}\nu_{i}}^\dagger +\text{arg}\mathcal{K}^{\llll}_{pq t_{0}\nu_{i}}\Big)\\
    \Delta \psi&= \text{arg}\mathcal{G}_{p\llll t_{i}\nu_{1}}  +\text{arg}\mathcal{G}_{q\llll t_{i}\nu_{1}}^\dagger+\text{arg}\mathcal{K}^{\llll}_{pq t_{i}\nu_{1}}\\
    &~~~~~~~~~~~~~~~~~~~~-\Big( \text{arg}\mathcal{G}_{p\llll t_{i}\nu_{0}}+\text{arg}\mathcal{G}_{q\llll t_{i}\nu_{0}}^\dagger +\text{arg}\mathcal{K}^{\llll}_{pq t_{i}\nu_{0}}\Big).   
\end{align*}

The average in Eq.~\ref{eq:ap}  always results in a net loss of amplitude for an off-phase centre source; in practice the  conditioning  $t_1-t_0 \rightarrow 0$ and $ \nu_1-\nu_0 \rightarrow 0$  cannot be satisfied because  the phase terms  in $\mathcal{K}^{\llll}_{pq t\nu}, \mathcal{G}_{p\llll t\nu}$ and $\mathcal{G}_{q\llll t\nu}^\dagger$  vary over time and frequency. 
This loss of amplitude is known as time and frequency decorrelation (or smearing) when it is caused only by the scalar $\mathcal{K}^{\llll}_{pq t\nu}$.  The general term  \textit{decoherence} is used when the amplitude loss is caused  by the  cumulative phase variation of $\mathcal{K}^{\llll}_{pq t\nu},  \mathcal{G}_{p\llll t\nu}, \mathcal{G}_{q\llll t\nu}^\dagger$ since $\mathcal{G}_{p\llll t\nu}$ and $\mathcal{G}_{q\llll t\nu}^\dagger$ also hold complex phases which  vary in time and frequency.  
Eq.~\ref{eq:ap} shows that for a full sky RIME, the decoherence can be measured separately as the sum of the contribution of each individual source. When considering the decoherence introduced by all the complex terms,  an approximation to the average of each individual source that is part of the sum in  Eq.~\ref{eq:ap} can be expressed as the phase differences. In the decoherence formulation discussed above,  we have neglected the case of decoherence caused by $\II_{\llll}$ because this work focuses on the decoherence of point sources rather than extended sources for which the phase term in $\II_{\llll}$ varies considerably over time and frequency. 

\begin{table}
\centering
\begin{tabular}{|p{1.2cm}|p{7cm} |} 
 \hline
 \textbf{Notation} & \textbf{Description} \\ [0.5ex] 
 \hline\hline
 $\Vtm$ & The data matrix for $N_{b}$ baselines before compression  \\ 
  $\Vtm_{n}$& The data matrix for $N_{b}$ baselines after compression\\
 $\VV$ & The unsampled visibility data  \\
 $\VVV_{pqt\nu}$ & The sampled visibility data for $pq$ at $t, \nu$  \\
 $\uu_{pq}$ & The baseline $pq$ vector \\
 $\II_{\llll}$ & The sky brightness in the direction of $\llll$\\
 $\Vtm_{pq}^{}$& The  rank$-r$ data matrix for baseline $pq$\\
 $\Vtm_{pq,n}$& The rank$-n$ approximation of  $\Vtm_{pq}^{}$ \\
 $\eta_{pqk}$ & The singular value\\
 $CF$   &The  compression factor\\
 $N_b$    & The number of baselines\\
 $M,N$     & The size of matrices\\
 $\epsilon$      & The maximum  threshold error \\
  $\varepsilon$         &  The  minimum  percentage signal to  preserve\\
   $\I^\mathrm{d}_{\mathrm{pq}}$    &  The dirty image for baseline $pq$\\
 $\widetilde{\I}^\mathrm{d}_{\mathrm{pq}}$    & The compressed  dirty image for baseline $pq$\\
 $\I^\mathrm{d}$          & The dirty image matrix\\
  $\widetilde{\I}^\mathrm{d}$    &The  compressed dirty image matrix\\
 $\I_{\mathrm{loss}}^\mathrm{d}$ & The loss image matrix\\
 $\FF$ & The Fourier transform operator\\
 $S/R$ &  The signal to noise\\
 \hline
\end{tabular}
\caption{Main mathematical notations: lowercase bold letters are vectors and uppercase bold letters are matrices. Calligraphic capital letters are used to designate functional forms. Everything else is a constant.}
\label{table:x}
\end{table}
\subsection{BDA and implementation}
\label{sec:implementation}
To aggressively compress visibility data  and mitigate amplitude lost caused by traditional averaging, several authors have discussed the potential of using BDA~\citep{cotton2009effects, atemkeng2018baseline, wijnholds2018baseline}. As shown in Eq.~\ref{eq:ap} the product of the three complex terms $\mathcal{K}^{\llll}_{pq t\nu}, \mathcal{G}_{p\llll t\nu}$ and $\mathcal{G}_{q\llll t\nu}^\dagger$ is attenuated by $sinc$ functions. The degree of attenuation at the edges of the FoV is determined by the width of these $sinc$ functions which depend on the spatial frequencies that each baseline samples over $[t_0, t_1]\times [\nu_0, \nu_1]$. On long spacing,  traditional averaging with a wide $[t_0, t_1]\times [\nu_0, \nu_1]$ will result in a narrow $sinc$ and significant drop in source amplitude. In order to keep amplitude loss equal across all baselines for a given FoV, the widths of the $sinc$ functions must remain constant, so that the time-frequency interval over which the data is averaged varies. This method is substantial for dense-core interferometric arrays where data is aggressively averaged on shorter baselines.  A small time-frequency interval is required to reach the width limit of the $sinc$ functions on longer baselines, resulting in less compression. 


It is advantageous for a BDA implementation to integrate with existing software, formats and specifications to benefit from compression, and for testing with real observational data.
The contemporary specification for radio astronomy data is the MS v2, based on the CASA Table Data System (CTDS) format.
We have, therefore, developed a BDA implementation targeting this specification and format.

MS rows are grouped by baseline, sorted by time and aggregated into bins whose
$\text{sinc}(\Delta \Phi/2)$ does not exceed the acceptable decorrelation tolerance defined by $\mathcal{K}^{\llll}_{pq t\nu}$.
Each bin's $\text{sinc}(\Delta \Phi/2)$  is inverted, firstly to obtain $\text{sinc}(\Delta \psi / 2)$ and secondly, the bin frequency integration interval $[\nu_0, \nu_1]$.
Practically, each bin can have a different frequency interval and this channelisation does not conveniently fit into the CTDS format which has a fixed number of channels per spectral window.

Therefore, to fit BDA data into this format, we discretise the channelisation by subdividing the original spectral window by the integral factors of its channels to produce new spectral windows representing each discretisation.
Then, each BDA  bin is output to a single row in the output MS and each row is associated with a different spectral window, thereby trading a small factor of compression for compatibility with the MS.

\subsection{Low-rank approximation: SVD}
\label{lowrankD}
The SVD break down a given complex data matrix $\Vtm$ having $M$ rows and $N$ columns into three independent matrices:
 \begin{align}
     \Vtm = \AAAA \LAMBDA \CC^\dagger,\label{eq:svd1}
 \end{align}
 where  $\LAMBDA $ is a diagonal matrix of size $M\times N$, $\AAAA$ of size $M\times M$ and $\CC$ of size $N \times N$ are unitary matrices. These matrices are defined  as:
   \begin{align}
\LAMBDA=\begin{bmatrix}
    \LAMBDA_r & \OO\\
    \OO & \OO 
\end{bmatrix}, ~~
\AAAA=[\AAAA_r, \aaa_{r+1}, \cdots, \aaa_{M}], ~~
\CC=\begin{bmatrix}
    \CC_{r}\\
    \cc_{r+1}\\
    \vdots\\
    \cc_{N}
   \end{bmatrix}.
   \end{align}
 The decomposition does not require $\Vtm$ to be a square matrix. Matrix $\LAMBDA_r=\mathrm{diag}(\eta_1, \eta_2, \cdots, \eta_r)$  of size $r\times r$ is diagonal with  $r=\min(M,N)$  and $\eta_k$ are the singular values of $\Vtm$. 
 As discussed in \citet{stewart1998matrix},  $\eta_k=\sqrt{\lambda_k}$  where $\lambda_k$ are the eigenvalues of $\Vtm^{\dagger}\Vtm$. We note that all the entries in the diagonal of $\LAMBDA_r$ are all non-zeros and are in decreasing order. Here, $\AAAA_r=[\aaa_1, \aaa_2, \cdots, \aaa_{r}]$ and $\CC_{r}=[\cc_1, \cc_2, \cdots, \cc_{r}]^\mathrm{T}$, where $^\mathrm{T}$ is the transpose operator.
 The vectors, $\aaa_k$ of size $M$ and $\cc_k$ of size $N$ are the left and right singular vectors of $\Vtm$, respectively.   $\aaa_k$ and $\cc_k$ are eigenvectors of $\Vtm \Vtm^{\dagger}$ and $\Vtm^{\dagger}\Vtm$, respectively. 
Eq.~\ref{eq:svd1} remains valid if rewritten as:
  \begin{align}
     \Vtm &=\AAAA_r \LAMBDA_r \CC_r^\dagger\\
     &=\sum_{k=1}^{r}\eta_k\aaa_k\cc_k^{\dagger}. \label{eq:svd2}
 \end{align}
 Each $\eta_k$ quantifies how best the corresponding component $\eta_k\aaa_k\cc_k^{\dagger}$ contributes to the relevant features in  Eq.~\ref{eq:svd2}; such as in the reconstruction of $\Vtm$.  The larger $\eta_k$   is, the more $\eta_k\aaa_k\cc_k^{\dagger}$ effectively contributes to the reconstruction of $\Vtm$. Note that $\aaa_k$ and $\cc_k^{\dagger}$ determine the geometry of the features contained in $\eta_k\aaa_k\cc_k^{\dagger}$. 

\section{Proposed lossy compression methods}
\label{methods}
The low-rank approximation methods described in this section are applied to the raw visibility data (continuous data), and at each baseline separately. We note that the methods can also be applied to gridded visibility data. However, the gridded data lie on a regular grid where data for all baselines have been interpolated together, making it difficult to gauge acceptable thresholding of the nonzero singular values. Also, since each baseline observes a different portion of the sky, the signal distortion and attenuation are baseline-dependent and the noise variance is different per-visibility data. These effects cannot be taken into account in gridded visibility data.  It is insufficient to store the images (therefore gridded visibility data) because typically calibration problems dominate and data has, out of necessity, be recalibrated most of the time to create science ready products for specific continuum, transient and line subdomains. This cannot be  done well with all the baselines being convolved onto grid spacing's.

Following the above recommendations, we formally express the visibility data (for a given baseline) in a compact and robust matrices formalism. Assume a non-polarized sky, a single channel timeslot visibility measured by a baseline $pq$ is the complex Stokes value:
\begin{align}
\VVV_{pqt_i\nu_j}
                 &= \delta_{pqij}\big(\VV + \mathcal{E}\big),\label{visandnoise} 
\end{align}
where $\delta_{pqij}=\delta(\uu-\uu_{pqt_i\nu_j})$ is a delta-function shifted to the $uv$-point being sampled, $\VV$ is discussed in Eq.~\ref{eq1} and $\mathcal{E}$ is the contaminating random noise with zero mean and r.m.s. $\Sigma$.  
 Assuming an observation of $M$ timeslots and $N$ channels, we can package the  visibility data for single baseline $pq$ into a single data matrix, $\Vtm_{pq}^{}$ of size $ M\times N$:
\begin{equation}
\Vtm_{pq}^{}=
  \begin{bmatrix}
    \VVV_{pqt_1\nu_1}^{} &\cdots& \VVV_{pqt_1 \nu_{N}} \\
    \vdots &  \cdots &\vdots \\
    \VVV_{pqt_{M} \nu_1}^{} & \cdots & \VVV_{pqt_{M} \nu_{N}}\\
  \end{bmatrix}.\label{resample1}
\end{equation}
The formalism can be extended to data structure with orthogonal bases capable of  fully describing the sky. 
We discuss the compression of $\Vtm_{pq}^{}$ and then, in Section~\ref{Effectsnoise}, we provide a detailed analysis of the compression  effect on $\Sigma$ and the sky image.  
Note that it is possible to remodel  $\Vtm_{pq}^{}$ into sub-matrices and independently find the low-rank of each sub-matrix. The latter  should be considered in cases where computation is a bottleneck (i.e. $MN \rightarrow \infty$) and where approximating local low-rank matrices is an advantage~\citep{lee2016llorma}.

\subsection{Method 1: \textit{Simple SVD}}
\label{sect:simpleSVD}
 Finding a very small low-rank data matrix of $\Vtm_{pq}$   without information loss is challenging because  the entries in $\Vtm_{pq}$ are strongly correlated.  Uncorrelated $\Vtm_{pq}$ is rarely possible in a real observation due to system electronics, and the fact that each visibility data point comes from a pairwise correlation between antenna voltages, etc. This implies that there exists no low-rank approximation, $\Vtm_{pq,n}$ of rank$-n$ that can perfectly reconstruct $\Vtm_{pq}$ of rank$-r$:
 \begin{align}
  \Vtm_{pq,n} &\simeq \Vtm_{pq}, ~ 1\leq n<r.
\end{align}
 \begin{align}
     \Vtm_{pq,n} &= \mathbf{A}_{pq,n}\mathbf{\Lambda}_{pq,n}\mathbf{C}^\dagger_{pq,n}\label{eq:simpleSVD1}\\
     &=\sum_{k=1}^{n}\eta_{pqk}\aaa_{pqk}\cc_{pqk}^{\dagger}.\label{eq:simpleSVD2}
 \end{align}
For simplicity, Eq.~\ref{eq:simpleSVD1} can also be written as a Kronecker product, $\otimes$ after vectorisation:
  \begin{align}
     \mbox{vec}(\Vtm_{pq,n}) &= (\mathbf{C}^\dagger_{pq,n}\otimes\mathbf{A}_{pq,n})\etapq_{pq},
 \end{align}
where $\etapq_{pq}=[\eta_{pq1}, \eta_{pq2}, \cdots, \eta_{pqn}]^\mathrm{T}$.

This \textit{simple SVD}  only allows an equal compression factor across all the baselines as $n$ is fixed.  The compression factor is the ratio between the number of data points of the uncompressed data and the number of data points of the compressed data. The entries of $\mathbf{A}_{pq,n}$ and $\mathbf{C}^\dagger_{pq,n}$ are complex numbers, while  the entries of $\mathbf{\Lambda}_{pq,n}$ are real numbers. In terms of computer storage requirements, if a complex entry counts as one entry, a real entry should count as $0.5$. Therefore, if the size of $\Vtm_{pq}$ is $M \times N$ then one can show that the number of elements in the sub-matrices needed to compute $\Vtm_{pq,n}$ is $n(M+N+0.5)$ which leads to an overall compression factor, $CF$ of:
\begin{equation}
    CF = \frac{MN}{n(M+N+0.5)}\label{eq:CFsimpleSVD}.
\end{equation}
 In this setting, one has to choose  $CF$ carefully to determine the number of singular values to retain on all the baselines:
\begin{equation}
        n = \ceil[\big]{\frac{MN}{CF(M+N+0.5)}},
\end{equation}
where $\ceil{.}$ is the ceiling operator. The compression loss is computed from  the tensor norm:
 \begin{align}
||\Vtm_{} - \Vtm_{n}||&\leq\epsilon ||\Vtm_{}||,\label{eq:epsilon}
 \end{align}
 where $\epsilon$ is the maximum  threshold error provided to control the divergence between the visibility data tensors $\Vtm_{}$ and $\Vtm_{n}$  representing the data for the $N_{b}$ baselines before and after compression,  respectively. In this work, we defined a tensor norm as the sum of the Frobenius norm, $||.||_{\mathrm{F}}$ per baseline: 
 \begin{align}
||\Vtm_{} - \Vtm_{n}||&=
    \sum_{pq}||\Vtm_{pq} - \Vtm_{pq,n}||_{\mathrm{F}}.\label{eq:v-vn}
 \end{align}
 Eq.~\ref{eq:v-vn} can be computed from the sum of the Euclidean norm of the singular values not retained at each baseline:
  \begin{align}
||\Vtm_{} - \Vtm_{n}||   &=\sum_{pq}\sqrt{\sum_{k=n+1}^{r}\eta_{pqk}^2}.\label{eq:v-vnF}
 \end{align}
Note that in Eq~\ref{eq:v-vnF}, $k$  runs from $n+1$ to $r$ i.e. $(r-n)$ singular values are not retained in the compression. $ ||\Vtm_{}||$ is defined as:
 \begin{align}
||\Vtm_{}||&=\sum_{pq}||\Vtm_{pq}||_{\mathrm{F}}\\
    &= \sum_{pq}\sqrt{\sum_{k=1}^{r}\eta_{pqk}^2}.
 \end{align}
  Instead of using a threshold error, a minimum  percentage of the signal to  preserve,  $\varepsilon$ can be specified: 
 \begin{align}
|| \Vtm_{n}||   &\geq\varepsilon||\Vtm_{}||,\label{eq:varepsilon}
 \end{align}
  \begin{align}
|| \Vtm_{n}||&=\sum_{pq}||\Vtm_{pq,n}||_{\mathrm{F}}\\
    &=\sum_{pq}\sqrt{\sum_{k=1}^{n}\eta_{pqk}^2}.
 \end{align}
If the constraint $\epsilon$ or $\varepsilon$ is given then it is computationally cheap to obtain the corresponding value of $n$ that satisfies  the constraint as shown in Algorithm~\ref{algo:1}. Note that these norms could directly be obtained from computing the Euclidean norm of $\mbox{vec}(\Vtm_{pq,n})$, $\mbox{vec}(\Vtm_{})$ and $\mbox{vec}(\Vtm_{n})$; the vectorisation versions of $\Vtm_{pq,n}, \Vtm_{}$ and $\Vtm_{n}$, respectively.

\begin{algorithm} 
 \KwData{$\Vtm_{}; \epsilon ~\text{or}~ \varepsilon$}
 \KwResult{the number of singular values to retain, $n$}
 $ n\leftarrow 0; \Vtm_{pq,n}\leftarrow \OO$\\
 \eIf{$\epsilon$ is given}{
 \Do{$\frac{\sum_{pq}||\Vtm_{pq} - \Vtm_{pq,n}||_{\mathrm{F}}}{\sum_{pq}||\Vtm_{pq}||_{\mathrm{F}}}>\epsilon$ and  $n\leq r$}{
 \For{all baseline $pq$}{
 $\Vtm_{pq, n}\leftarrow \Vtm_{pq,n-1}+\eta_{pqn}\aaa_{pqn}\cc_{pqn}^{\dagger}$\\
 }
  $n\leftarrow n+1$
}}{
 \Do{$\frac{\sum_{pq}||\Vtm_{pq,n}||_{\mathrm{F}}}{\sum_{pq}||\Vtm_{pq}||_{\mathrm{F}}}\times 100<\varepsilon$ and $n \leq r$}{
 \For{all baseline $pq$}{
 $\Vtm_{pq, n}\leftarrow \Vtm_{pq,n-1}+\eta_{pqn}\aaa_{pqn}\cc_{pqn}^{\dagger}$\\
 }
  $n\leftarrow n+1$
}}
 \caption{Finding $n$ for \textit{simple SVD}. Here, all the $\Vtm_{pq}$ are taken from $\Vtm_{}$}\label{algo:1}
\end{algorithm}

\subsection{Source position in the sky and baselines effects on the distribution of singular values}
\label{sect:skyposition}
As discussed in Section~\ref{RIMEaveraging} the baselines observe different portions of the sky
due to the variation of the phase terms in $\mathcal{K}^{\llll}_{pq t\nu}, \mathcal{G}_{p\llll t\nu}$ and $\mathcal{G}_{q\llll t\nu}^\dagger$, as well as the different Nyquist sampling due to the baselines geometry and distribution. With this information, we noted that decorrelation is baseline-dependent. Given this finding, the following questions arise: \textit{Is the distribution of the singular values equal across all baselines?} If not, \textit{how do the strength and decay of the singular values differ at each baseline? Can the variation in strength and decay of the singular values be used to aggressively compress the data and what limits the compression?}  

\begin{figure*}
\centering
  \includegraphics[width=.95\linewidth]{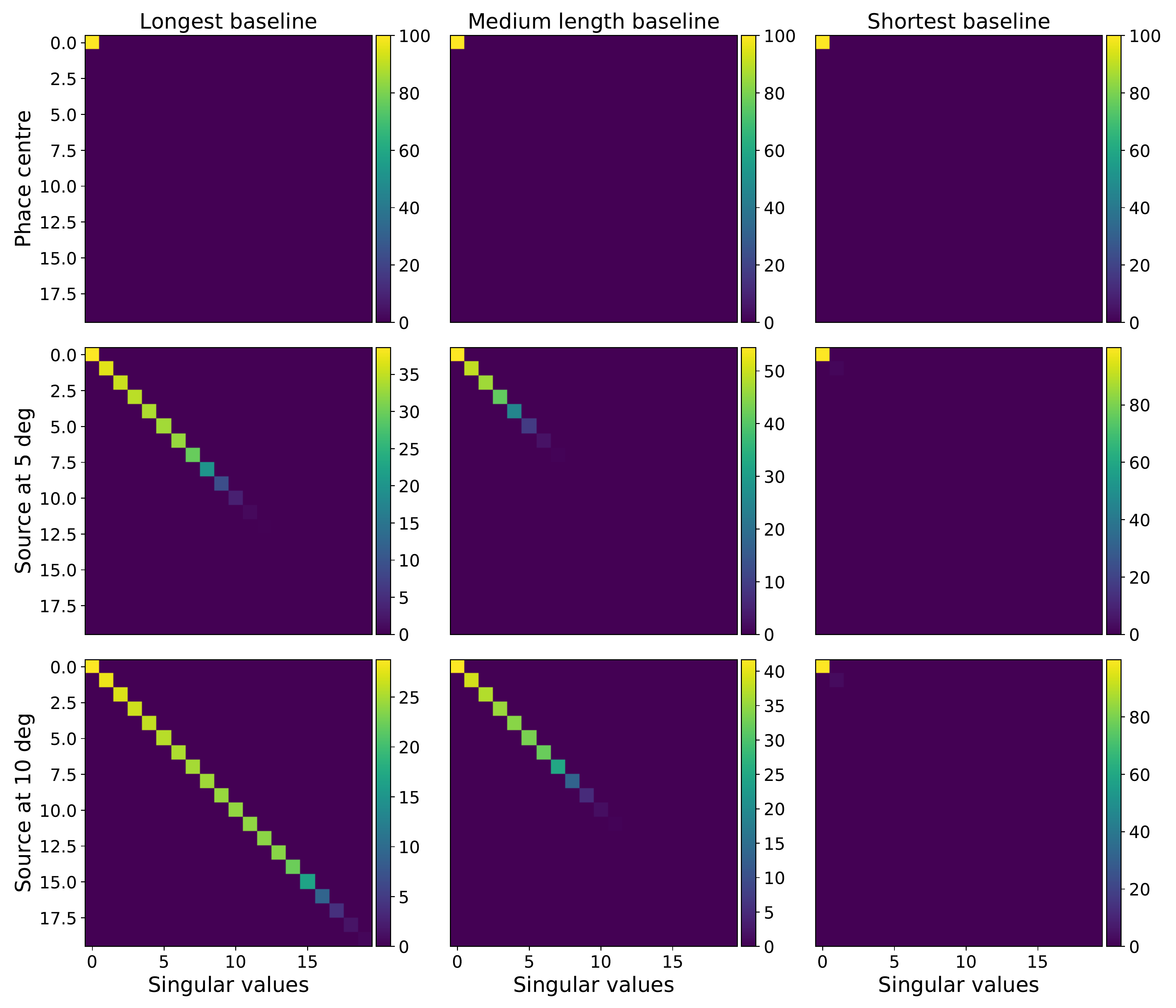}
\caption{The first 20 singular values of three baselines; the longest baseline (left-column panels), medium-length baseline (middle-column panels) and shortest (right-column panels) of a 1 Jy source at the phase centre (top-row panels), 5 deg (middle-row panels) and 10 deg (bottom-row panels) away from the phase centre. The singular values are obtained from simulating the MeerKAT telescope at  1.4 GHz. The data is sampled at 1 s and 80 kHz during $166$ min $40$ s with 0.8 MHz bandwidth.}
\label{fig:1}
\end{figure*}

\begin{figure*}
\centering
  \includegraphics[width=.3\linewidth]{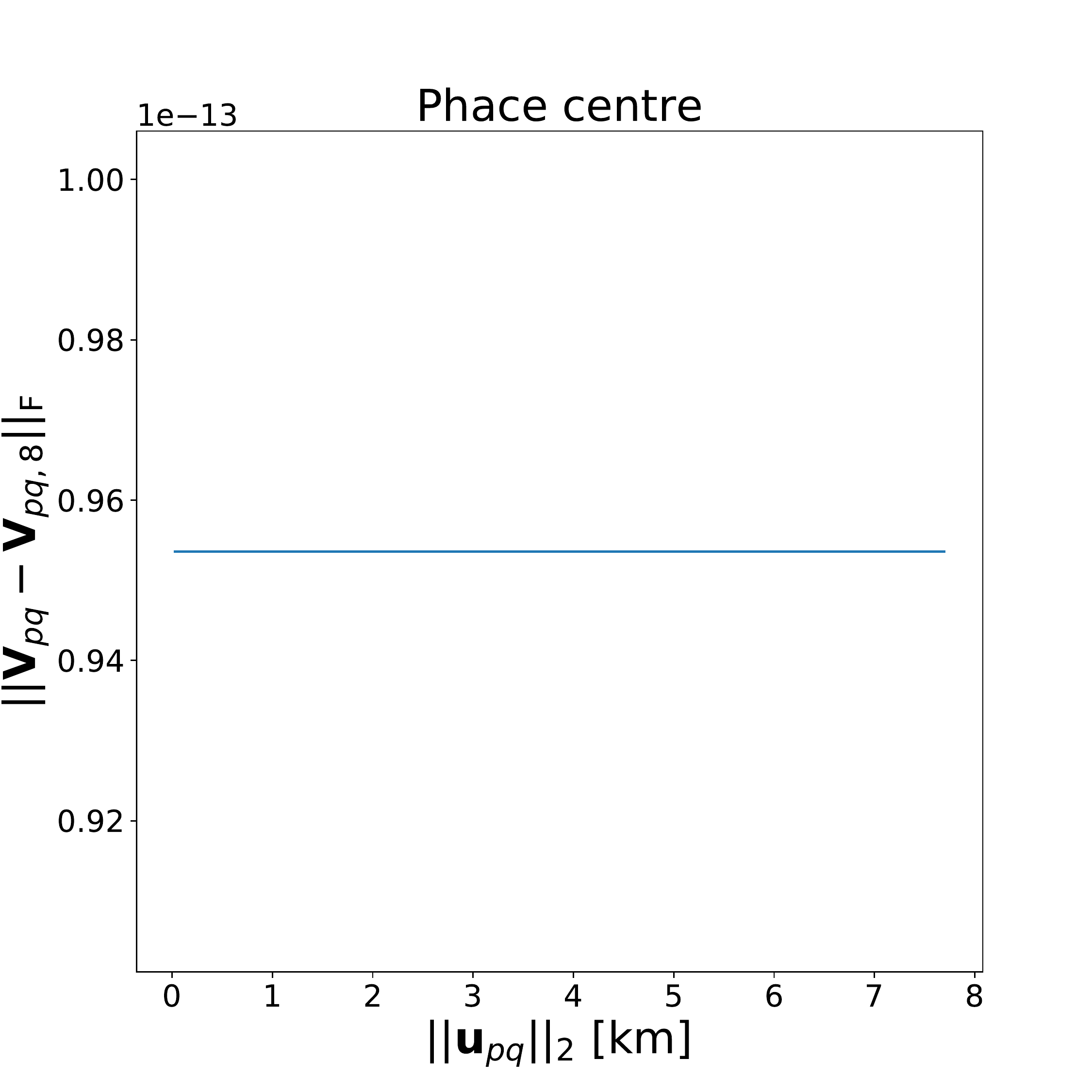}
   \includegraphics[width=.3\linewidth]{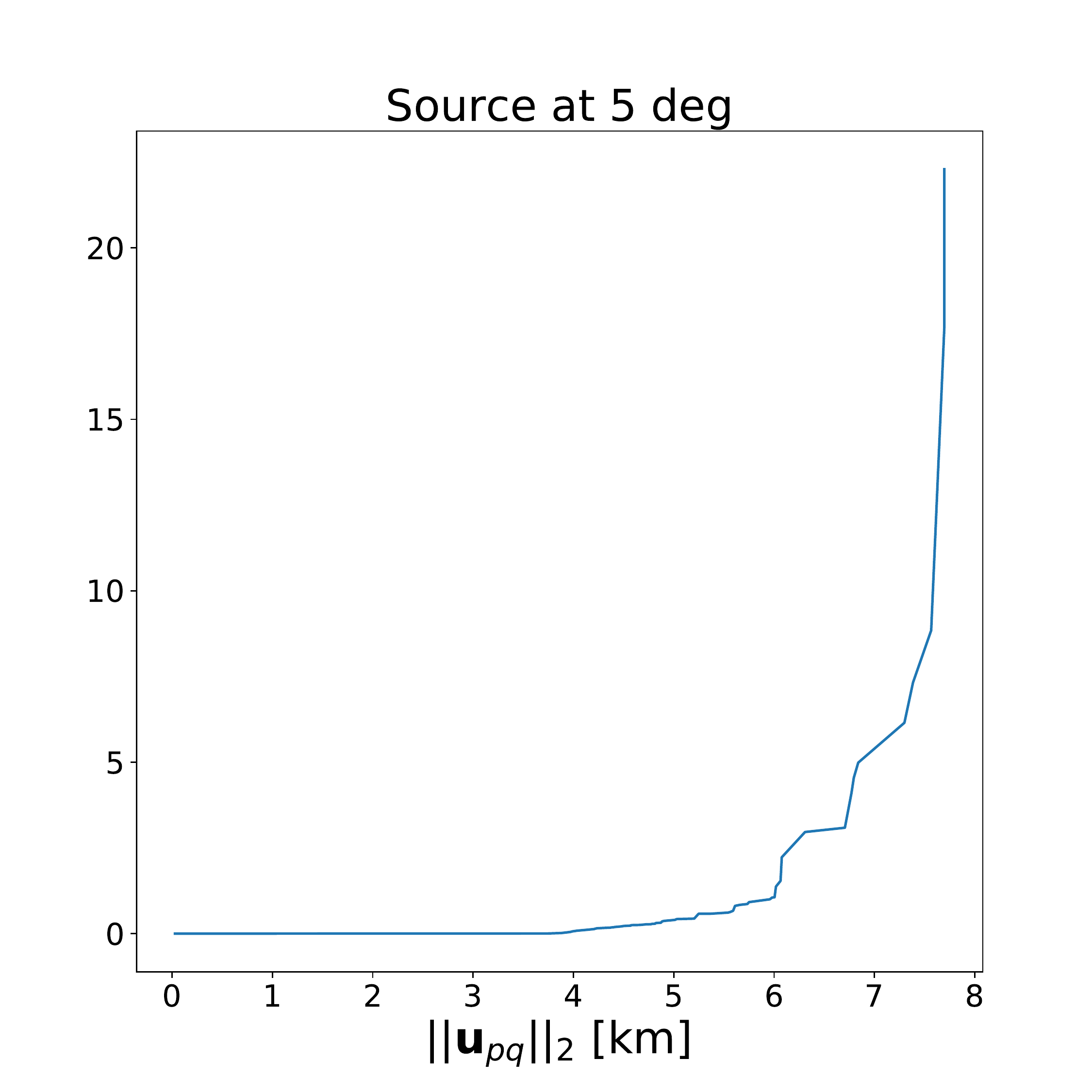}
    \includegraphics[width=.3\linewidth]{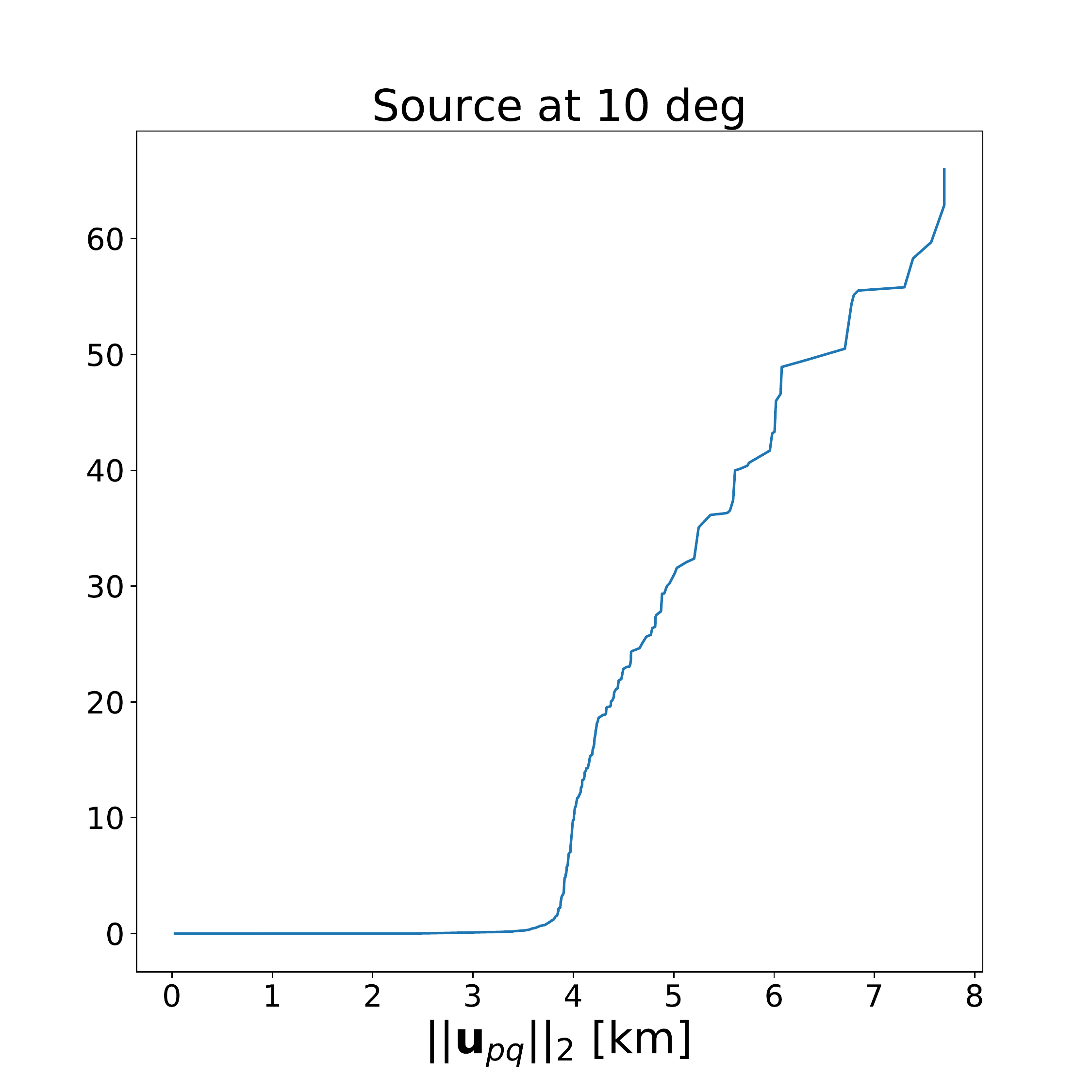}
\caption{Compression errors from the simulation in Figure~\ref{fig:1}, where the first eight  components of each of the baselines are retained. At the phase centre, the compression error tends to zero for all baselines. On long baselines,  the compression error increased rapidly for far-field sources. The compression error is small on short baselines compared to long baselines.}
\label{fig:xxx}
\end{figure*}

To answer the above questions, Figure~\ref{fig:1} shows the first $20$ singular values of three baselines; the longest baseline (left-column panels), medium-length baseline (middle-column panels) and shortest (right-column panels) of a 1 Jy source at the phase centre (top-row panels), 5 deg (middle-row panels) and 10 deg (bottom-row panels) away from the phase centre. The singular values are obtained by simulating the MeerKAT telescope at  1.4 GHz. The simulation corresponds to $10~000$ timeslots per second with a total bandwidth of 0.8 MHz divided into 10 channels each of width 80 kHz. This figure shows some important behaviors of the distribution of the singular values across baselines, therefore,  it deserves a detailed explanation. Regardless of the position of the source, the first component in each baseline captures the most features in the original data. In addition, the strength and decay of the singular values are different at each baseline and source position, with shorter baselines decaying rapidly compared to longer baselines where the decay is slow. At the phase centre of the observation, the singular values decay faster; the degree of decay becomes slower as the source moves away from the phase centre. In other words:
\begin{itemize}
\item  For a source at the phase centre, the first component in all the baselines captures the entire features in the original data, in other words   $\eta_{pq1}$ is very large and $\forall k >1, \eta_{pqk}\sim 0$. This result aligns with what is known; for all baselines, decorrelation is negligible at the phase centre of the observation. 
    \item  The spread of the singular values is baseline-dependent; the singular values are spread to several components on the longer baselines while for shorter baselines only a few of the first components capture features; for example, say $pq$ and $\alpha\beta$ are two baselines with length $||\textbf{u}_{pq}||_2<||\textbf{u}_{\alpha\beta}||_2;\exists i$ such that $\forall k>i$:
 \begin{align}
\eta_{pqi}&\gg \eta_{\alpha\beta i} ~\text{and}~\eta_{\alpha\beta k}\gg \eta_{pqk},
 \end{align}
 where $||.||_2$ is the Euclidean norm.
This result aligns with the fact that longer baselines observed severe amplitude loss compared to  short baselines.
    \item  The spread of the singular values is position-dependent; the singular values are spread in many components for far-field sources compared to nearby field sources. This is easy to explain - far-field sources experience severe amplitude loss compared to nearby field sources.
\end{itemize}
 The above results show that if we were to consider the same number of  $n$ components to retain (i.e. $r-n$ components to discard) across all the baselines as described in Section~\ref{sect:simpleSVD}, this will have a direct impact on baseline compression errors, $||\Vtm_{pq} -\Vtm_{pq, n}||_{\mathrm{F}}$, which will be small on short  baselines compared to long baselines. Figure~\ref{fig:xxx} shows these compression errors from the simulation in Figure~\ref{fig:1}, where the first eigh components of each of the baselines are retained. In this figure, the compression errors are plotted against baseline lengths; left-, middle-, and right panels are the compression errors when the source is at the phase centre, $5$ deg and $10$ deg,  respectively.  The following is observed:
 \begin{itemize}
 \item   At the phase centre, $||\Vtm_{pq} -\Vtm_{pq, 8}||_{\mathrm{F}}\sim 0$ for all baselines $pq$.
  \item On longer baselines, $||\Vtm_{pq} -\Vtm_{pq, 8}||_{\mathrm{F}}$ increased rapidly for far-field sources.  
 \item For any  two baselines of mismatched length, say $pq$ and $\alpha\beta$ with $||\textbf{u}_{pq}||_2<||\textbf{u}_{\alpha\beta}||_2$ we have:
 \begin{align}
||\Vtm_{pq} -\Vtm_{pq,8}||_{\mathrm{F}}<||\Vtm_{\alpha\beta} -\Vtm_{\alpha\beta,8}||_{\mathrm{F}}.\label{eq:errorsel}
 \end{align}
 \end{itemize}
 We note that the singular values experience a slow decay for  out-of-phase centre sources and on long baselines. Since the strength  of a singular value, $\eta_{pqk}$ indicates the degree of features that the component, $\eta_{pqk}\aaa_{pqk}\cc_{pqk}^{\dagger}$ contributes to the original data for baseline $pq$, depending on how the singular values are decaying for the  given baseline,  the components with smaller  singular values could be discarded. By so doing, we observe the following:
\begin{itemize}
\item A significant number of components could be discarded on shorter baselines with little or no loss of source amplitude. This leads to aggressive data compression with negligible effect on image fidelity.
\item  While attempting to preserve the loss of source amplitude and image fidelity, only a few components could be discarded on longer baselines.
\end{itemize}
 This makes the compression baseline-dependent. However, 
  for a given FoV, a sophisticated compression approach is to carefully choose $n$ on each of the baselines so that the errors in Eq.~\ref{eq:errorsel} are equal on all the baselines.  This method is described in Section~\ref{bdsvd}.
\subsection{Method 2: Baseline-dependent SVD (\textit{BDSVD})}
\label{bdsvd}
The \textit{BDSVD} method finds the corresponding baseline-dependent number of components to retain on each baseline so that the compression errors between baselines do not vary, for example for any  two baselines $pq$ and $\alpha\beta$ with $||\textbf{u}_{pq}||_2< ||\textbf{u}_{\alpha\beta}||_2$ find the  number of components $n_{pq}$ and $n_{\alpha\beta}$, respectively so that:
 \begin{align}
||\Vtm_{pq} -\Vtm_{pq,n_{pq}}||_{\mathrm{F}}\equiv ||\Vtm_{\alpha\beta} -\Vtm_{\alpha\beta,n_{\alpha\beta}}||_{\mathrm{F}},
 \end{align}
 where 
 \begin{align}
     \Vtm_{pq,n_{pq}} 
     &=\sum_{k=1}^{n_{pq}}\eta_{pqk}\aaa_{pqk}\cc_{pqk}^{\dagger}.\label{eq:simpleSVD}
 \end{align}
  It should be noted that  the maximum  threshold error $\epsilon$ (or  minimum  percentage of the signal to  preserve  $\varepsilon$) is now verified on each individual baseline as opposed to Eq.~\ref{eq:epsilon} (or to Eq.~\ref{eq:varepsilon}): 
 \begin{align}
||\Vtm_{pq} - \Vtm_{pq,n_{pq}}||_{\mathrm{F}}&\leq\epsilon ||\Vtm_{pq}||_{\mathrm{F}}
 \end{align}
  \begin{align}
|| \Vtm_{pq, n_{pq}}||_{\mathrm{F}}  \geq\varepsilon||\Vtm_{pq}||_{\mathrm{F}}.
 \end{align}
Intuitively, since $||\textbf{u}_{pq}||_2< ||\textbf{u}_{\alpha\beta}||_2$ this means that $n_{pq}<n_{\alpha\beta}$ while the cumulative strength of the singular values of a compact source are equal on all baselines, 
\begin{align}
\sum_{k=1}^{n_{pq}}\eta_{pqk}^2=\sum_{k=1}^{n_{\alpha\beta}}\eta_{\alpha\beta k}^2. \label{eq:samenumretainedvariance}
 \end{align}
With this information, Eq.~\ref{eq:same} is correct and valid.
\begin{align}
|| \Vtm_{n}||
    &=\sum_{pq}\sqrt{\sum_{k=1}^{n_{pq}}\eta_{pqk}^2}\\
    &=N_{b}\sqrt{\sum_{k=1}^{n_{pq}}\eta_{pqk}^2}.\label{eq:same}
 \end{align}
 In the above, the number of components to retain is baseline-dependent, thus the compression factor also becomes baseline-dependent:
  \begin{equation}
    CF_{pq} = \frac{MN}{n_{pq}(M+N+0.5)}\label{eq:CFpq},
\end{equation}
which is larger for shorter baselines compared to longer baselines, for example  $CF_{pq}>CF_{\alpha\beta}$ since $n_{pq}<n_{\alpha\beta}$.
As opposed to Eq.~\ref{eq:CFsimpleSVD}, the \textit{BDSVD} overall $CF$ follows:
 \begin{equation}
    CF = \frac{N_{b}MN}{(M+N+0.5)\sum_{pq} n_{pq}}. \label{eq:CFsBDSVD}
\end{equation}
  Algorithm~\ref{algo:2} describes an iterative process to find  the baseline-dependent number of component, $n_{pq}$ given $\epsilon$ or $\varepsilon$. 
  
For compression factor $CF$, the space-saving $SS$ is measured as:
  \begin{equation}
   SS=(1-CF^{-1})\times 100 \%\label{eq:SSsBDSVD}.
\end{equation}

 \subsection{Computation complexity and parallelisation}
Low-rank approximation has computational drawbacks. It is expensive to compute the SVD of a very large matrix. For visibility data compression as discussed in this paper, for some cases (e.g. to archive the data), we can overlook this computational drawback since we have to compress the data only once at a time. In other cases, for example big data radio interferometers such as the SKA, the amount of visibility data on each of the baselines is very large, even the once-off compression can become a challenge since finding the exact $CF_{pq}$  involves the computation of singular values at each baseline sequentially, which is an expensive computational task. For example, if the visibility data for each baseline has size $M\times N$ with $M\geq N$, then for $N_b$ baselines, the sequential full rank$-n$ \textit{simple SVD} scales as:
\begin{equation}
  s_\mathrm{cost}\sim  \mathcal{O}\big(N_bMN^2\big). \label{eq:Seqcost}
\end{equation}
\textit{BDSVD} scales closely with \textit{simple SVD} for equal  $CF$.
However, the good news is that visibility data has some natural basis consisting of row data.  This information can be used to speed up the compression via a parallel algorithm. The row visibility data for a given baseline can be shared with multiple compute nodes; for example,  for each baseline, we can subdivide the row data into chunks of $\Vtm_{pq}^\mathrm{[i]}$ and compute the SVD for all the 
 chunks in parallel. The subscript $^{[i]}$ indicates the $i^{th}$ chunk of the row visibility data. We briefly discuss this distributed compression process in Algorithm~\ref{algo:3}, however, further investigation is needed to assess its computational efficiency in practice and to determine at what rate of visibility data size the algorithm should be enforced.

Algorithm~\ref{algo:3} shows that we can compute the SVD for each of the baselines in parallel (line 1); assume $N_p$ parallel nodes. Also (as shown in line 5) for each baseline, to find $\mathbf{A}_{pq},\mathbf{\Lambda}_{pq}$ and $\mathbf{C}_{pq}$ involve computing all  $\Vtm_{pq}^\mathrm{[i]\dagger}\Vtm_{pq}^{[i]}$ in parallel.  $\Vtm_{pq}^\dagger\Vtm_{pq}$ is then  obtained from summing all $\Vtm_{pq}^\mathrm{[i]\dagger}\Vtm_{pq}^{[i]}$. 
The latter suggests that the complexity of Algorithm~\ref{algo:3} scales as:
\begin{align}
  p_\mathrm{cost}&\sim  \mathcal{O}\bigg(N_bMN^2/N_p\bigg).
\end{align}
Ideally, we would hope to demonstrate strong scaling over Eq.~\ref{eq:Seqcost}, but this remains outside the scope of this work.

\section{Effects on the image and noise: analytical quantification}
\label{Effectsnoise}
The SVD is not perfectly orthogonal in practice, both noise and  signal are present in each of the components of the decomposition. A trade-off between the amount of signal and noise to remove is necessary. \textit{Simple SVD}
and \textit{BDSVD} also result in a small loss of signal and retain a small amount of noise. Our goal now is to provide the mathematical models  to quantify the signal loss and the noise removed as a form of contribution  at all baselines.  

\subsection{Effect on the image}
As the Fourier transform $\FF$ is linear, the compressed  dirty image $\widetilde{\I}^\mathrm{d}_{\mathrm{pq}}$  of a single baseline is the sum:
\begin{align}
     \widetilde{\I}^\mathrm{d}_{\mathrm{pq}}&=\FF\Vtm_{pq,n_{pq}} \\&=\sum_{k=1}^{n_{pq}}\eta_{pqk}\I_{pqk}^\mathrm{d},\label{eq:kthimage}
 \end{align}
  where $\eta_{pqk}\I_{pqk}^\mathrm{d}$   is the $k^{th}$ linearly independent component of the sky image seen by the baseline $pq$ and $\FF$ is a unitary linear operator. In relation to Parseval’s theorem,  we have:
  \begin{align}
   \I_{pqk}^\mathrm{d}=\mathbf{a}_{pqk}^{\prime}\mathbf{c}^{\prime\dagger}_{pqk},\label{eq:FFTSingularVector}
  \end{align}
  where $\mathbf{a}_{pqk}^{\prime}$ and $\mathbf{c}^{\prime}_{pqk}$ are vectors.  The singular value, $\eta_{pqk}$ indicates how best $\eta_{pqk}\I_{pqk}^\mathrm{d}$ contributes to the quality and fidelity of $\widetilde{\I}^\mathrm{d}_{\mathrm{pq}}$. A closer look at Eq.~\ref{eq:kthimage} shows that the singular values in both the visibility and image domains are equal, thanks to the linear and unitary properties of the Fourier transform. This means that the choice of the domain where the data should be compressed does not matter;  the number of components to be retained in the visibility domain would eventually be the same in the image domain.
   \begin{algorithm} 
 \KwData{$\Vtm_{}; \epsilon ~\text{or}~ \varepsilon$}
 \KwResult{the list of the number of singular values to retain at each baseline $\big\{n_{pq}\big\}, \forall pq$}
 $ n_{pq}\leftarrow 0; \Vtm_{pq,n_{pq}}\leftarrow \OO$\\
 \eIf{$\epsilon$ is given}{
 \For{all baseline $pq$}{

 \Do{$\frac{||\Vtm_{pq} - \Vtm_{pq,n_{pq}}||_{\mathrm{F}}}{||\Vtm_{pq}||_{\mathrm{F}}}>\epsilon$ and $n_{pq}\leq r$}{
 $\Vtm_{pq, n_{pq}}\leftarrow \Vtm_{pq,n_{pq}-1}$\\
 $~~~~~~~~~~~~~~~~~~~~~~+\eta_{pqn_{pq}}\aaa_{pqn_{pq}}\cc_{pqn_{pq}}^{\dagger}$
}
$n_{pq}\leftarrow n_{pq}+1$\\
}}{
 \For{all baseline $pq$}{
 \Do{$\frac{|| \Vtm_{pq,n_{pq}}||_{\mathrm{F}}}{||\Vtm_{pq}||_{\mathrm{F}}}\times 100<\varepsilon$ and $n_{pq}\leq r$}{
 $\Vtm_{pq, n_{pq}}\leftarrow \Vtm_{pq,n_{pq}-1}$\\
 $~~~~~~~~~~~~~~~~~~~~~~+\eta_{pqn_{pq}}\aaa_{pqn_{pq}}\cc_{pqn_{pq}}^{\dagger}$
}
 $n_{pq}\leftarrow n_{pq}+1$\\
}
 }
 \caption{Finding $\big\{n_{pq}\big\}, \forall pq$ using \textit{BDSVD}. Here, all the $\Vtm_{pq}$ are taken from $\Vtm_{}$.}\label{algo:2}
\end{algorithm}

  Alternatively, the dirty image  is derived from summing Eq.~\ref{eq:kthimage} across all the baselines:
  \begin{align}
     \widetilde{\I}^\mathrm{d}&=\sum_{pq} \widetilde{\I}^\mathrm{d}_{\mathrm{pq}}\\
     &=\sum_{pq}\bigg(\sum_{k=1}^{n_{pq}}\eta_{pqk}\I_{pqk}^\mathrm{d}\bigg).
 \end{align}
If $\I_\mathrm{}^\mathrm{d}=\sum_{pq}\FF\Vtm_{pq}$ is the Fourier transform of the  uncompressed data (i.e. the uncompressed image of the sky), then to quantify the net loss in signal  per-pixel, $\I_{\mathrm{loss}}^{\mathrm{d}}$, the following difference is  adopted as a  standard fidelity metric: 
\begin{align}
    \I_{\mathrm{loss}}^\mathrm{d}&=\I^\mathrm{d}-\widetilde{\I}^\mathrm{d}\\
                    &=  \sum_{pq}\bigg(\sum_{k=1}^{r}\eta_{pqk}\I_{pqk}^\mathrm{d}-\sum_{k=1}^{n_{pq}}\eta_{pqk}\I_{pqk}^\mathrm{d}\bigg).
\end{align}
For $n_{pq}< r$, we have
\begin{align}
    \I_{\mathrm{loss}}^\mathrm{d}                  &=\sum_{pq}\sum_{k=n_{pq}+1}^{r}\eta_{pqk}\I_{pqk}^\mathrm{d},
    \end{align}
where the entries of $\I_{\mathrm{loss}}^\mathrm{d}$ are different from $0$.
\begin{algorithm} 
 \KwData{$\Vtm_{}$; $N_{p}$}
 \KwResult{The SVD of $\Vtm_{}$ decomposed per baseline: $\big\{\mathbf{A}_{pq},\mathbf{\Lambda}_{pq},\mathbf{C}_{pq}\big\}, ~\forall pq$}
 \For{all baseline $pq$, in parallel}{
    \For{all $i$, in parallel}{
         $\Vtm_{pq}^\dagger\Vtm_{pq} += \Vtm_{pq}^\mathrm{[i]\dagger}\Vtm_{pq}^\mathrm{[i]}$
            }
    
    $\mathbf{C}_{pq}\mathbf{\Lambda}_{pq}^2\mathbf{C}_{pq}^\dagger=\Vtm_{pq}^\dagger\Vtm_{pq}$\\
    $\mathbf{\Lambda}_{pq}=\sqrt{\mathbf{\Lambda}_{pq}^2}$\\
     \For{all $i$, in parallel}{
         $\mathbf{A}_{pq}^\mathrm{[i]}=\Vtm_{pq}^\mathrm{[i]}\mathbf{C}_{pq}\mathbf{\Lambda}_{pq}^{-1}$
            }
          $\mathbf{A}_{pq}=\big\{\mathbf{A}_{pq}^\mathrm{[i]}\big\}, ~\forall i$  
}
 \caption{Finding $\big\{\mathbf{A}_{pq},\mathbf{\Lambda}_{pq},\mathbf{C}_{pq}\big\}, \forall pq$ using parallel computing. Here, all the $\Vtm_{pq}$ are taken from $\Vtm_{}$, and $N_p$ is the number of compute nodes.}\label{algo:3}
\end{algorithm}
\subsection{Noise filtering  and S/N}
\label{noise:filtering}
Three important questions arise surrounding noise filtering when  \textit{simple SVD} and \textit{BDSVD} are used i) \textit{how does the noise behave for  \textit{simple SVD} where the same number of components are retained on all the baselines? ii) how does \textit{BDSVD} and its varying compression factor affect the noise at each baseline and how does the filtered noise differ from the \textit{simple SVD} and  traditional averaging? and iii) on what type of baseline is noise heavily filtered when using \textit{simple SVD} and   \textit{BDSVD}?} In this section, we address these questions through well-posed conditioning and discussion. 

The entries of the compact visibility data matrix $\Vtm_{pq}^{}$ come from sampling the  sum of $\VV$  and noise $\mathcal{E}$ as shown in Eq.~\ref{visandnoise}. Thus, each $\eta_{pqk}$ reflects the  strength of the signal contribution from a component which is sampled from $\VV+\mathcal{E}$. Small $\eta_{pqk}$ corresponds to components of $\Vtm_{pq}$ that are heavily corrupted by noise, which means that if one retains components with the bigger $\eta_{pqk}$ this should be equivalent to removing noise and keeping useful signal. This means that with \textit{simple  SVD} and \textit{BDSVD}, the noise in the compressed data is reduced for $n<r$ and $n_{pq}<r$, respectively. Below, we provide details on the noise filtering capability of \textit{simple SVD} and \textit{BDSVD} compared to  traditional averaging at the same $CF$. 

 For the same compression factor, the analytical visibility noise penalty estimates of \textit{simple SVD} and $\textit{BDSVD}$ is the relative decrease in noise over traditional averaging:
  \begin{align}
      \Xi_X&=\frac{\Sigma_{X}}{\Sigma_{avg}}, \label{newm}
  \end{align}
  where $\Sigma_{X}$ and $\Sigma_{avg}$ are compressed noise  using $\textit{simple SVD}$ (or $\textit{BDSVD}$) and  traditional averaging,  respectively. The goal is to show that $\Xi_X<0$ which means $\Sigma_{X}<\Sigma_{avg}$.
  
  Assuming  that  for all baselines and samples, the noise term has constant r.m.s. $\Sigma$, when denoising a signal by averaging $n_{avg}$ samples, the reduction in noise is well understood to be  $\Sigma_{avg}=\Sigma/\sqrt{n_{avg}}$ if the noise is not correlated between samples; which confirms that $\Sigma_{avg}<\Sigma$. As discussed, the SVD removed some noise in the data, therefore, $\Sigma_{X}<\Sigma$. However, at the same compression factor, it is   not trivial to see that $\Sigma_{X}<\Sigma_{avg}$ analytically (we refer the reader to Appendix \ref{ap:A} for a detailed mathematical discussion).  Empirical measurements are  used in Section~\ref{simulation} to show that  at the same compression factor $\Sigma_{X}<\Sigma_{avg}$ and therefore  $ \Xi_X<0$.
   
   In the image domain, the noise penalty estimate of the centre pixel is given by
   \begin{align}
      \Xi_{X}^{\WW}&=\frac{\sum_{pqt_i\nu_j}\WW_{pqt_i\nu_j}^2\Xi_{X}^2}{(\sum_{pqt_i\nu_j}\WW_{pqt_i\nu_j})^2}, \label{gainimageplane}
  \end{align}
  where $W_{pqt_i\nu_j}$ is the imaging weight per visibility. 

  As discussed above, components with smaller $\eta_{pqk}$ are strongly noisy and are candidates for removal when noise is of concern. In addition, components with larger $\eta_{pqk}$ contain the most signal to retain. \textit{Simple SVD} and \textit{BDSVD} only retain components with larger $\eta_{pqk}$ for given thresholding and components that do not meet this threshold are rejected. \textit{BDSVD} differs from \textit{simple  SVD} in that the number of components,  $n_{pq}$ to retain is baseline-dependent. 
As mentioned in Section~\ref{sect:skyposition}, the  first $\eta_{pqk}$ are large for shorter baselines and decrease faster compared to longer baselines, where $\eta_{pqk}$  decreases slowly (Figure~\ref{fig:1} clearly shows this behaviour of $\eta_{pqk}$). This means that if the same number of components is retained on all the baselines as in \textit{simple SVD}, then some of the components that contain the signal will be discarded on longer baselines, hence distorting the compressed data. On the other hand,  on the shorter baselines,  several noisy components are retained in addition to all the components with a strong signal.  \textit{BDSVD} takes advantage of this drawback and constructs a baseline-dependent number of components to retain such that only components with strong signal strength are retained on each baseline. This heavily filters out noise and maintains signal fidelity compared to \textit{simple SVD} at the same $CF$. With this in mind, then at the same compression factor, we can write:
  \begin{align}
      \Sigma_{\mathrm{svd}}&>\Sigma_\mathrm{bdsvd}\implies \Xi_{\mathrm{svd}}^{\WW}>\Xi_{\mathrm{bdsvd}}^{\WW},\label{ssvdbdsvd}
  \end{align}
where $\Sigma_\mathrm{svd}$ and $\Sigma_\mathrm{bdsvd}$  (respectively $\Xi_{\mathrm{svd}}^{\WW}$ and $\Xi_{\mathrm{bdsvd}}^{\WW}$) are the noise (respectively the noise penalty) using \textit{simple SVD} and \textit{BDSVD},   respectively. Analytically, Eq.~\ref{ssvdbdsvd} clearly shows that \textit{BDSVD} reduces noise compared to \textit{simple SVD}. Empirical measurements are  used in Section~\ref{simulation}  to confirm the analytical result in Eq.~\ref{ssvdbdsvd}.
 
 The metric we used to measure the signal to noise, $S/R$  in decibel in each pixel of the compressed image is:
   \begin{align}
     S/R&=10\log_{10}\frac{\widetilde{\II}_{\llll}}{\sigma_{\mathrm{pix}}+c_{\mathrm{noise}}},\label{SNR}
 \end{align}
 where $\widetilde{\II}_{\llll}$ is the compressed version of $\II_{\llll}$; the sky without any electronics corruption and effects that can disrupt the signal in the path towards the instrument (see Eq.~\ref{eq1}). In this formulation, $c_{\mathrm{noise}}$ represents the signal coming from sources that are outside the FoV and $\sigma_{\mathrm{pix}}$ is the per-pixel noise in the dirty image:
   \begin{align}
      \sigma_{\mathrm{pix}}^2&=\frac{\sum_{pqt_i\nu_j}\WW_{pqt_i\nu_j}^2\Sigma_X^2}{(\sum_{pqt_i\nu_j}\WW_{pqt_i\nu_j})^2}.
  \end{align}

\section{Simulations}
\label{simulation}
The MeerKAT and the The European Very Long Baseline Interferometry Network (EVN) are used as reference telescopes in this  section to evaluate and compare the performance of each method; traditional averaging,  BDA, \textit{simple SVD} and \textit{BDSVD}. Three different metrics are evaluated; i) source amplitude loss is measured vs. baseline lengths, baseline-dependent compression factor vs. baseline lengths and east-west baseline lengths, ii) amplitude loss is measured relative to the source's position in the sky; iii) to measure the spectral and temporal resolution, the Point Spread Function (PSF) shape is measured relative to the source's position in the sky  and (iv) the S/R is also measured relative to the source's position in the sky. 

Note that for a rank$-n$ SVD decomposition of  two-dimensional visibility data, $\Vtm_{pq,n}$ as defined in Eq.~\ref{eq:simpleSVD1}; $M$  and $N$ are the number of time steps and channels, respectively. In the space where $\Vtm_{pq,n}$ is defined, the row and column data are  the  time and frequency observations, respectively. When $\Vtm_{pq,n}$ is projected into the rank$-n$ SVD space, the time and frequency observations are mixed, which makes it difficult to completely determine the direction in which the  compression of rank$-n$ is performed.

To compare the proposed methods with  traditional averaging and BDA, we need to evaluate the above metrics in a precise compression direction, 
 such as in time or frequency. We adopt the following strategy, for example, to evaluate the metrics in time: $\Vtm_{pq,n}$ is scaled from a data matrix of size $M \times N$ to size $M_1 \times M_2 \times N$, where rank$-n$ SVD is only applied  in the time direction with   chunk of size $M=M_1 \times M_2$. The notation $CF=CF_t\times CF_\nu$ adopted in this section means that the data is compressed by a factor of $CF_t$ in time and $CF_\nu$ in frequency. The discussion in this section does not include the compression in the frequency direction ($CF_\nu=1$); however, equivalent performance is observed in the frequency direction if  $\Vtm_{pq,n}$ is scaled to $M \times N_1 \times N_2$ and the rank$-n$ SVD is applied only to entries $N=N_1  \times N_2$ of $\Vtm_{pq,n}$.

\subsection{MeerKAT telescope}
\label{MeerkATTel}
For the MeerKAT telescope at 1.4 GHz, a point source is simulated during  $166$ min $40$ s which gives $10~000$ time steps of 1 s each. The simulation has a total bandwidth of 0.8 MHz divided into 10 channels each 80 kHz wide.  The 1.4 GHz MeerKAT telescope observes a full FoV with radius $\sim 2.25$ deg; the second null of the PB without time and frequency smearing effects with a time step of 1 s and a channel width of 80 kHz. This high-resolution dataset is scaled to $100\times 100\times 10$, where \textit{simple SVD} and \textit{BDSVD} are applied separately at each frequency chunk of size $100\times 100$. The compression factors we adopt are $CF =25\times 1$, $CF= 35\times 1$ and $CF= 50\times 1$, which translates to a space-saving of $96.0\%$ , $97.14\%$ and $98.0 \%$ in the time direction, respectively. With this high-resolution dataset, $CF= 50\times 1$ is the maximum time compression factor that can be achieved with the proposed methods.  This is because $CF_t=50$ for rank-$1$ SVD; in other words only one component is kept across all baselines and the compression factor as described in Eq.~\ref {eq:CFpq} becomes $100\times 100/(100+100 +0.5)\sim 50$.

The strategies adopted for  traditional averaging  and BDA are:
 \begin{itemize}
     \item  Two low-resolution MSs with bins size $25s\times 5kHz$ and $50 s \times 5 kHz$ are created to receive the resampled visibilities for  traditional averaging with $CF=25\times 1$ and  $CF= 50\times 1$, respectively.
     \item Using the method described in Section~\ref{sec:implementation}, a third MS is created to receive the resampled visibility for BDA with  $CF= 25\times 1$ and $CF =50\times 1$. 
      \item A fourth MS of the same copy as the high-resolution  MS is created to receive resampled visibility for \textit{simple SVD} and \textit{BDSVD}. No backup policy  is required to save the decompressed visibility data for \textit{simple SVD} and \textit{BDSVD}. The adopted compression factors are $CF= 25\times 1$ and $CF= 50\times 1$ for \textit{simple SVD}, while for \textit{BDSVD} $CF=25\times 1$ and $CF=35\times 1$ are adopted. Since $CF=50\times 1$ is the maximum compression factor  \textit{BDSVD} with $CF=50\times 1$ will be equivalent to \textit{simple SVD} with $CF=50\times 1$. Due to the latter,  we do not run \textit{BDSVD} with $CF=50\times 1$.
 \end{itemize}

\subsubsection{Amplitude vs. east-west baseline lengths}
\label{amplVSEW}
\begin{figure}
\centering
  \includegraphics[width=1.\linewidth]{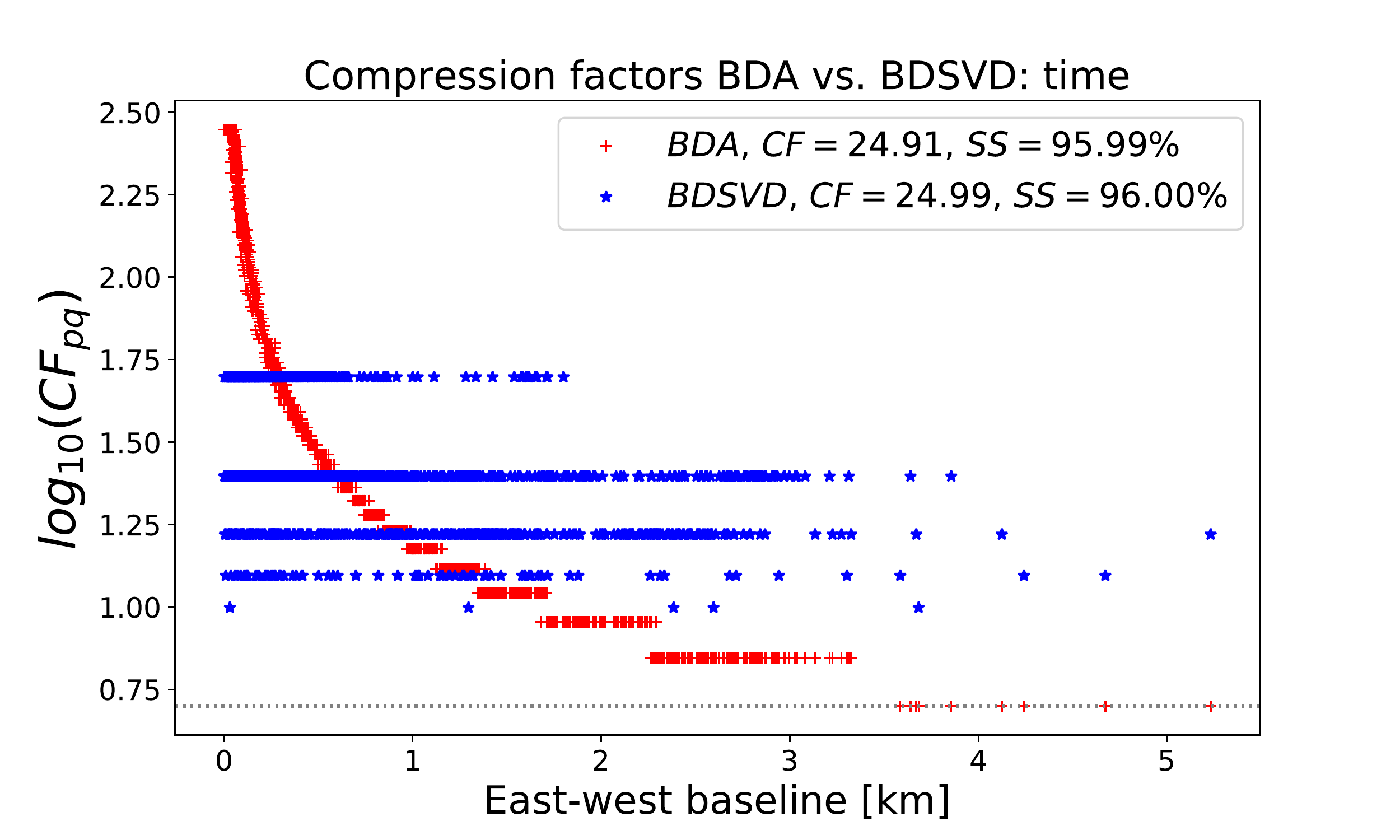}
    \includegraphics[width=1.\linewidth]{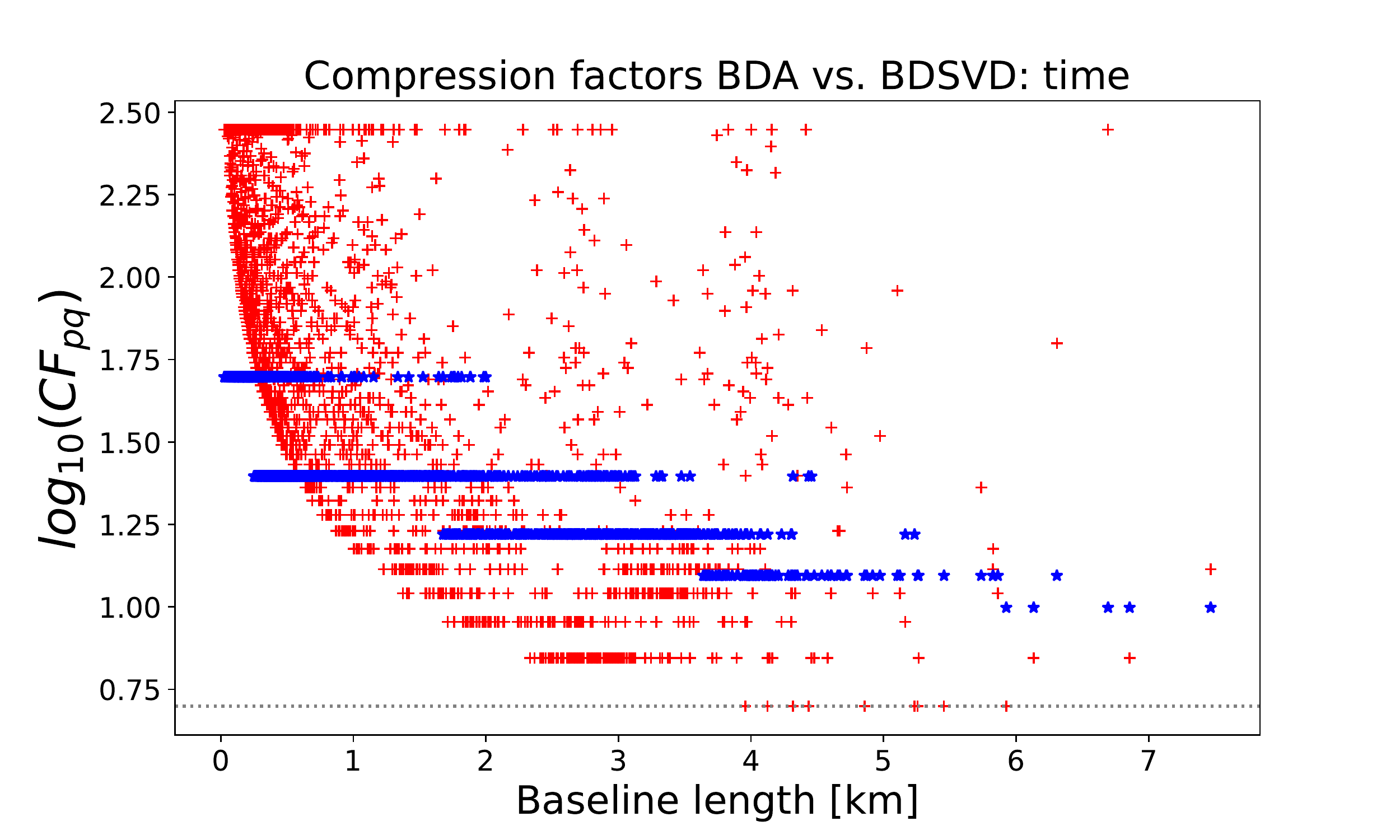}\\
       \includegraphics[width=1.\linewidth]{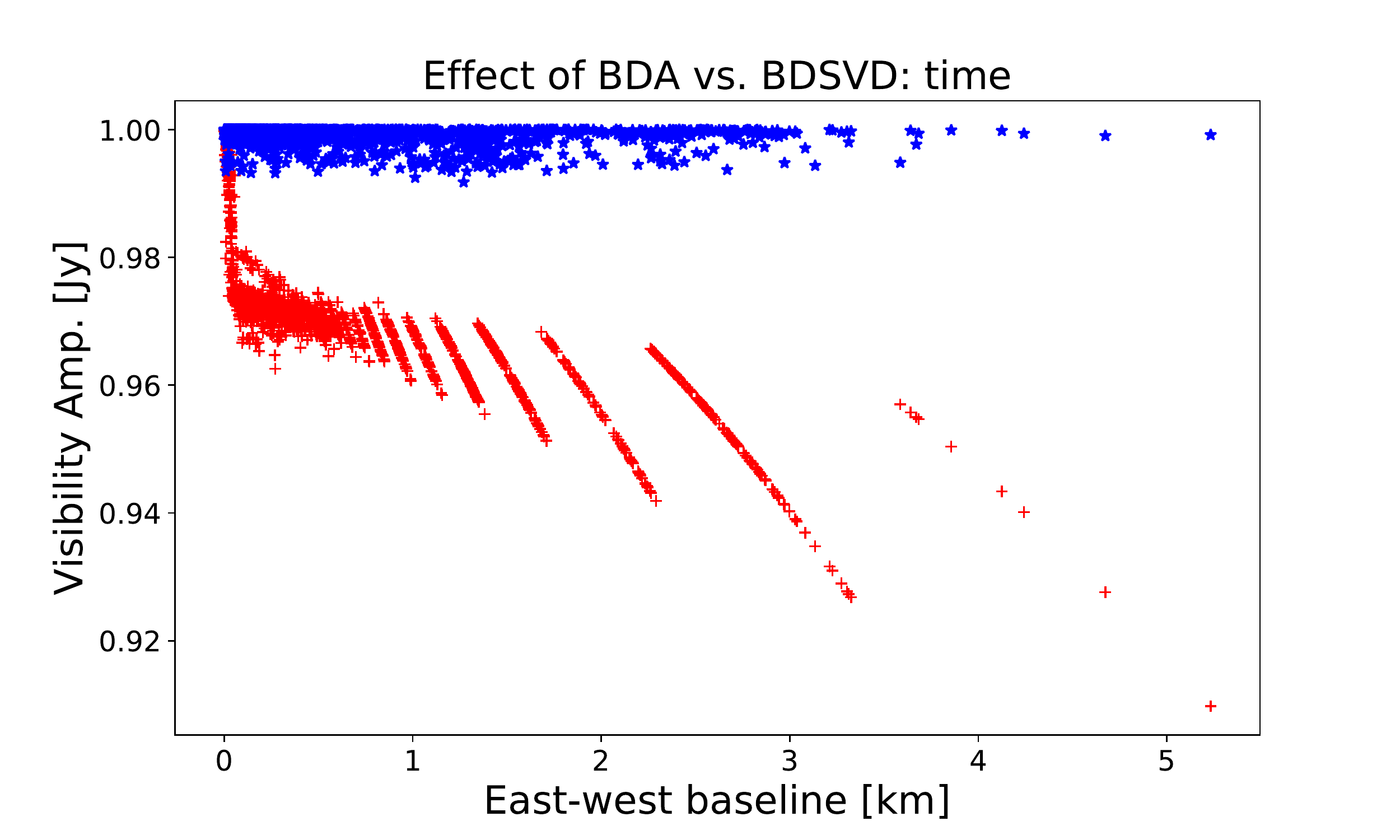}
\caption{The top panel shows the baseline-dependent compression factor $CF_{pq}$ in the logarithmic scale against
increasing east-west baseline lengths while  the middle panel  shows $CF_{pq}$ in logarithmic scale as a function of baseline length. The bottom panel shows the amplitude of the 1 Jy point source against  increasing east-west baseline lengths. BDA  sees  strictly different $CF_{pq}$ than \textit{BDSVD} while \textit{BDSVD} outperformed BDA in preserving the 1 Jy source amplitude.}
\label{fig:bda1source}
\end{figure}

This first test aims to quantify the  decorrelation of a single point source of  1 Jy amplitude placed at 2.25 deg; the second null of the MeerKAT at 1.4 GHz PB.  The results are shown in Figure~\ref{fig:bda1source}. 

For $\varepsilon=99\%$, Figure~\ref{fig:bda1source} (top panel)  shows the baseline-dependent compression factor $CF_{pq}$ in logarithmic scale against increasing east-west baseline lengths. The rank$-n$ SVD-related methods, $CF_{pq}$ is computed from Eq.\ref{eq:CFpq} after calculating all 
 $n_{pq}$ using Algorithm~\ref{algo:2}, while for BDA $CF_{pq}$ is computed following the discussion in \cite{atemkeng2018baseline}.
This is an important result of this study, so it deserves a detailed explanation. 

Although we understand the $CF_{pq}$ of BDA as discussed in \cite{atemkeng2018baseline}, the  $CF_{pq}$ of \textit{BDSVD} shows a different pattern and strength for east-west projected  baselines. It can be observed that for BDA the $CF_{pq}$ are strictly decreasing for increasing east-west projected baseline length, whereas for \textit{BDSVD}, east-west projections of equal lengths can have different $CF_{pq}$ where some  $CF_{pq}$ are significantly larger than others.  However, this behaviour still shows that more data is compressed on the smaller east-west baseline lengths compared to the longer; although the pattern and strength are not similar to that of BDA. Figure ~\ref{fig:bda1source} (middle panel) shows the $CF_{pq}$ as a function of increasing baseline length. We observe that baselines of equal lengths can have different $CF_{pq}$ when BDA or \textit{BDSVD} are in effect. Whereas $CF_{pq}$ shows more decreasing strengths with \textit{BDSVD} and randomly distributed with BDA, which confirms that decorrelation is not a degree of baseline length but rather the  east-west baseline length.   At the same overall compression factor $CF$, BDA and \textit{BDSVD} see  strictly different $CF_{pq}$. 

Figure ~\ref{fig:bda1source} (bottom panel) shows the source amplitude of 1 Jy as a function of increasing east-west projection lengths. It is clear from this result that \textit{BDSVD} outperformed  BDA. The source amplitude is attenuated with BDA compared to \textit{BDSVD} which retains more than  $\varepsilon=99\%$ of the source amplitude. Additionally, we observe  with \textit{BDSVD} that the amplitude of the decorrelated source is not constant for some equal east-west baseline lengths. Indeed, as observed in Figure ~\ref{fig:bda1source} (top panel); $CF_{pq}$ varies for certain equal east-west baseline lengths, which can result in different degrees of decorrelation. Similar behaviour can be observed with BDA, where decorrelation is revealed faster for east-west baseline lengths belonging to the same averaging   range. For example, all east-west baselines with a length between $2.2~km$ and $3.5~km$ fall within the same averaging range where $\sim 2$ bins are averaged together. Since the length of the east-west baseline in the same averaging range  is different; explaining the  different degrees of decorrelation.

\subsubsection{Amplitude vs. source position}
\label{sec:sourceamp}
This section examines the result of the amplitude-decorrelation with respect to the source position in the sky. The noise penalty is   also measured and compared for each of the compression methods. To measure the thermal noise, the high-resolution dataset as described above is used to populate an empty sky with 1 Jy thermal noise, where the different compression methods are applied.
\begin{figure*}
\centering
  \includegraphics[width=0.92\linewidth]{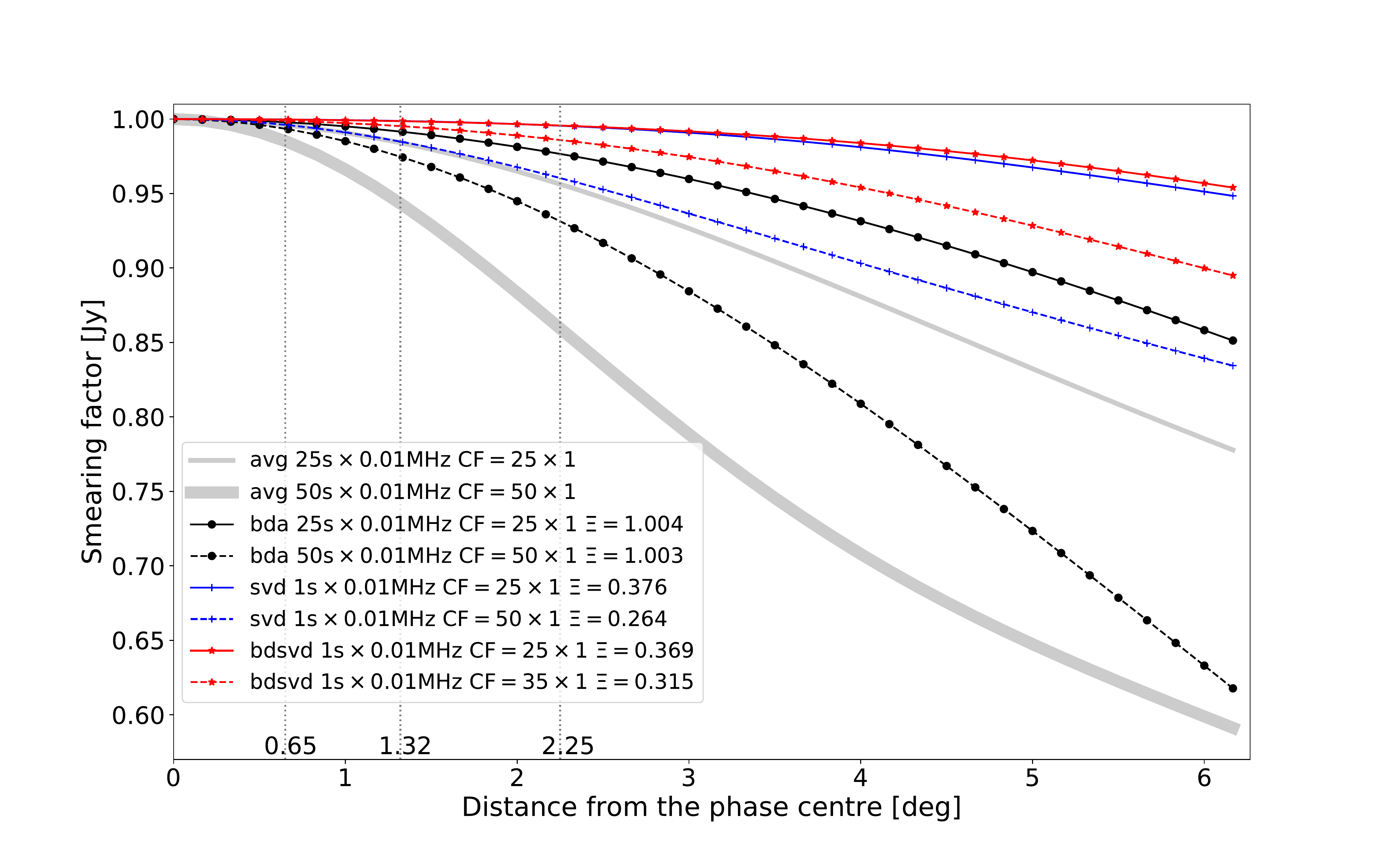}
\caption{MeerKAT telescope observing at 1.4 GHz during $166$ min $40$ s with a total bandwidth of 0.8 MHz; demonstrating the degree of a 1 Jy  source amplitude loss at various sky positions  and the associated noise penalty. Smearing against source distance from the phase centre, for  traditional averaging and BDA with $CF=25 \times 1$ and $CF=50 \times 1$ and for  \textit{simple SVD}, \textit{BDSVD} with $CF=25 \times 1$, $CF=35 \times 1$  and $CF=50\times 1$. The space-saving $SS$  and  the noise penalty $\Xi$ are given relative to the   traditional averaging bins.}
\label{fig:meerkattime}
\end{figure*}

Figure~\ref{fig:meerkattime} shows the results of time compression. There are important points to note about these results. A  compression factor of  $CF=25\times 1$  is needed to retain at least $\varepsilon=99\%$ of the source amplitude. At this compression regime, \textit{simple SVD} and \textit{BDSVD} keep the source amplitude almost flat with negligible  attenuation of $1\%$ that starts around a radius of $\sim 4$ deg  upwards, while the $1\%$ attenuation starts around a radius of $\sim 1.32$ deg and $\sim 0.8$ deg upward for the BDA and  traditional  averaging,  respectively. A compression factor of $CF=50\times 1$ provides up to a FoV with radius $\sim 2.9$ deg for $\varepsilon=95\%$ whereas at this compression factor, traditional averaging and BDA can only provide a FoV with radius up to $\sim 1.32$ deg and $\sim 2$ deg,  respectively.

 It can also be noted that  traditional averaging only provides a FoV with a radius of $\sim 1.32$ deg for $CF =25\times 1$ whereas  to obtain the same FoV with a radius of $\sim 1.32$ deg using \textit{simple SVD}, we can compress the data by two-order of magnitude higher than traditional  averaging. Traditional averaging or BDA cannot compress data by a factor of $CF =25\times 1$ and achieve $\sim 2.25$ deg radius with $1\%$ attenuation. However, at this $1\%$ attenuation rate, \textit{BDSVD} achieves $\sim 2.25$ deg radius for  $CF=35\times 1$.  \textit{BDSVD} draws its potential from the advantages of both BDA and \textit{simple  SVD}.
 
 The values of $\Xi\sim 1$ for BDA, this result has also been  observed in \cite{atemkeng2018baseline} and at the same compression factor $CF$, the reduction in noise of both BDA and traditional averaging varies with  $\sqrt{CF}$. Values of $\Xi< 1$ for \textit{simple SVD},  which demonstrates the common use of SVD to denoise signals. The values of $\Xi$ for \textit{BDSVD} are  lower than that of \textit{simple SVD} which confirms the theoretical result discussed in Section~\ref{noise:filtering};  the noise performance of \textit{BDSVD} exceeds that of \textit{simple SVD} at the same compression factor.
 
\subsubsection{PSF distortion vs. source position}
\begin{figure*}
\centering
  \includegraphics[width=.9\linewidth]{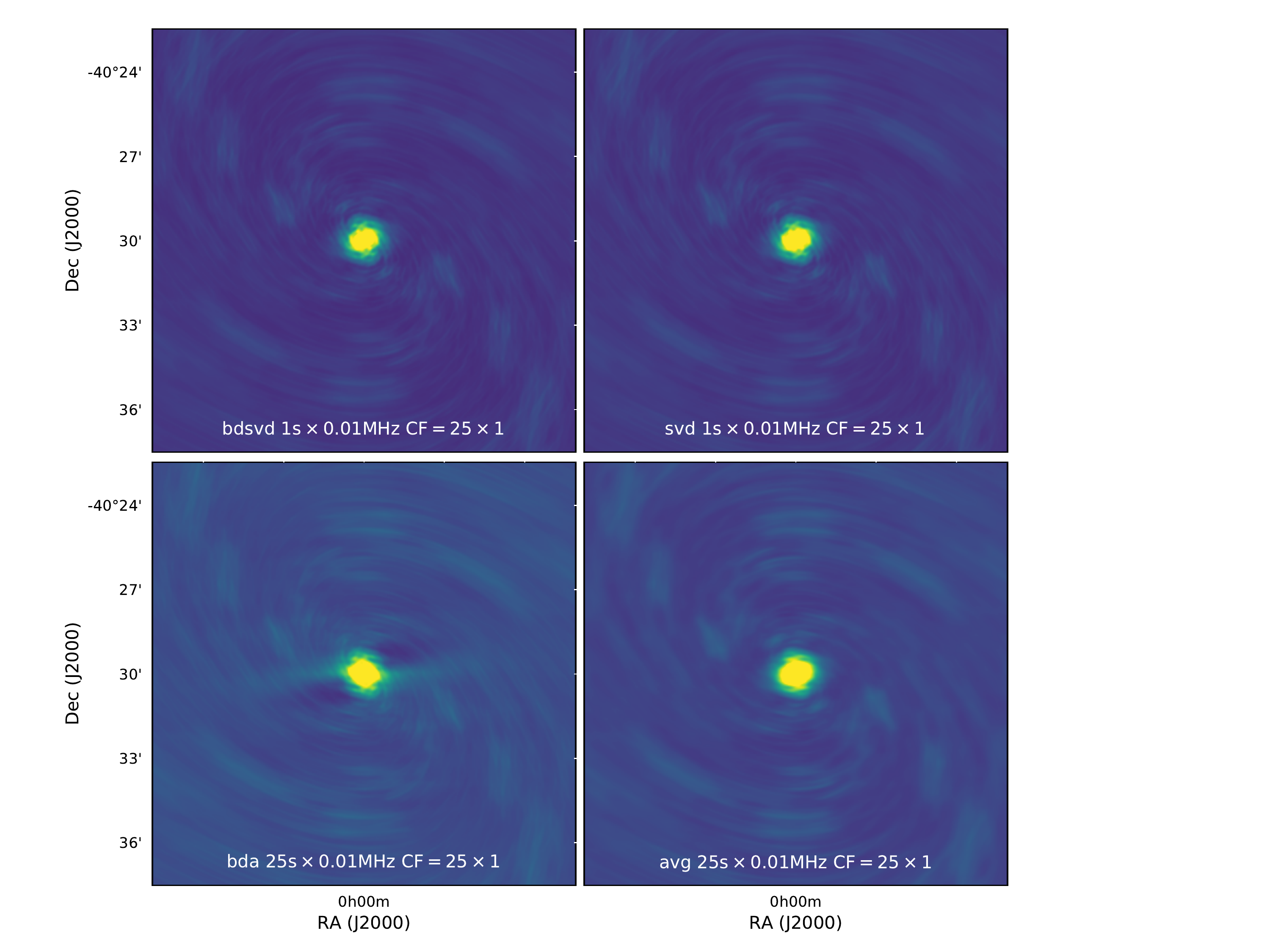}\\
  \includegraphics[width=.45\linewidth]{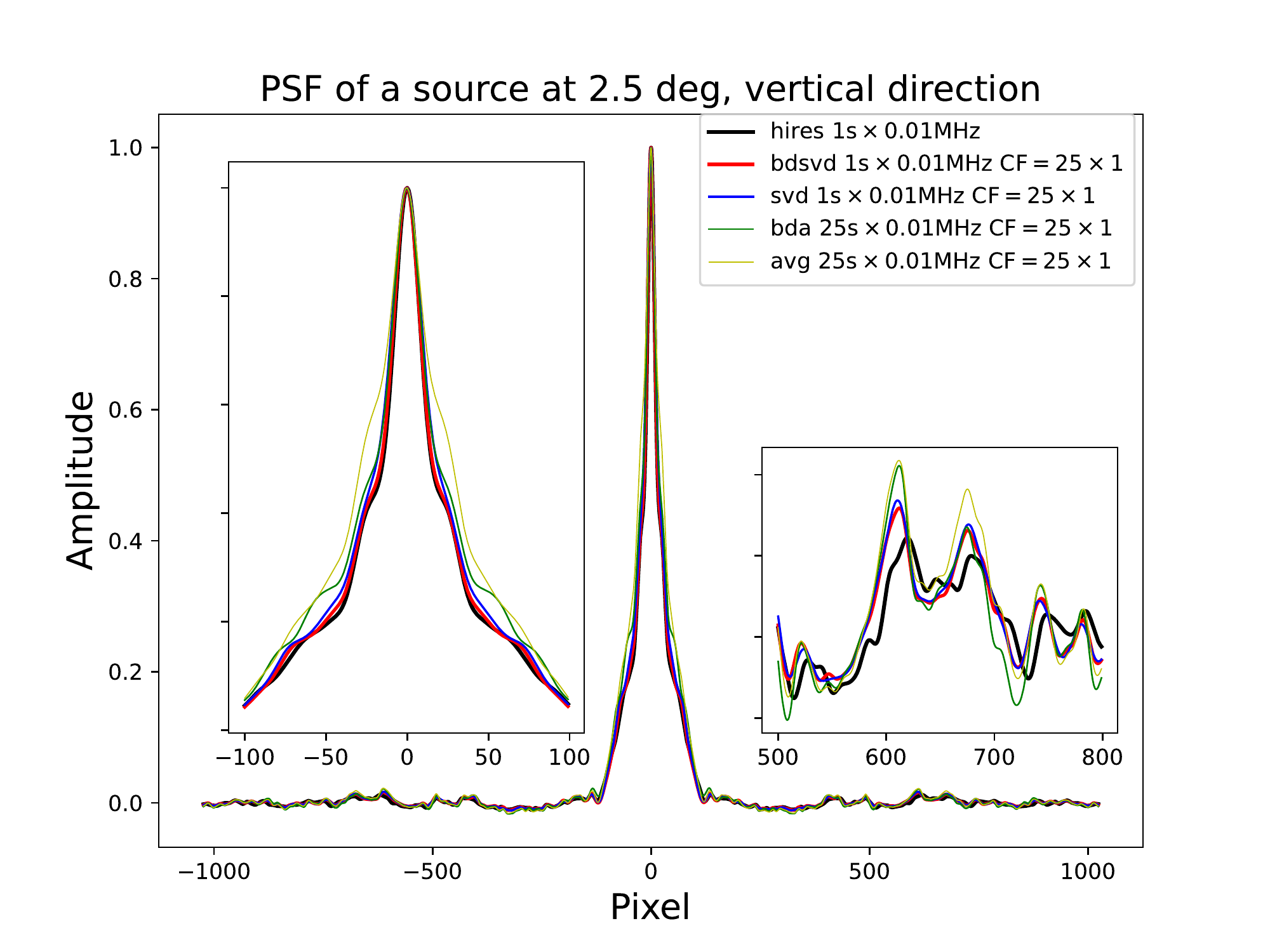}
  \includegraphics[width=.45\linewidth]{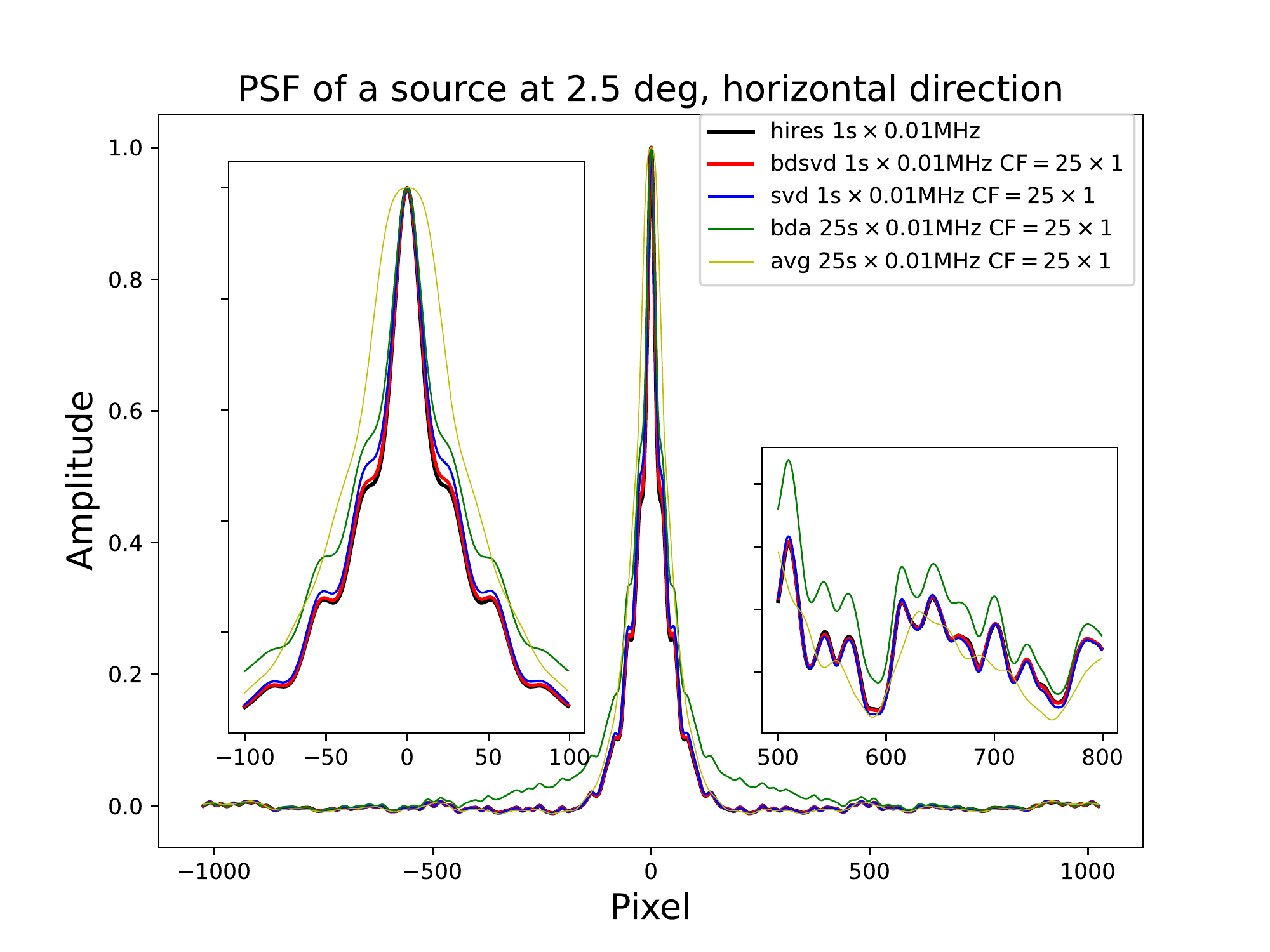}
\caption{MeerKAT telescope observing at 1.4 GHz during $166$ min $40$ s with a total bandwidth of 0.8 MHz. The PSF of a source at $2.25$ deg is shown when the high-resolution dataset is compressed for $CF=25\times 1$ with \textit{BDSVD} (top left), \textit{simple SVD} (top right), BDA (middle left),  traditional averaging (middle right), the vertical (bottom left) and horizontal (bottom right) directions of the normalised PSFs.}
\label{fig:psfdistortion1}
\end{figure*}
\begin{figure*}
\centering
  \includegraphics[width=0.92\linewidth]{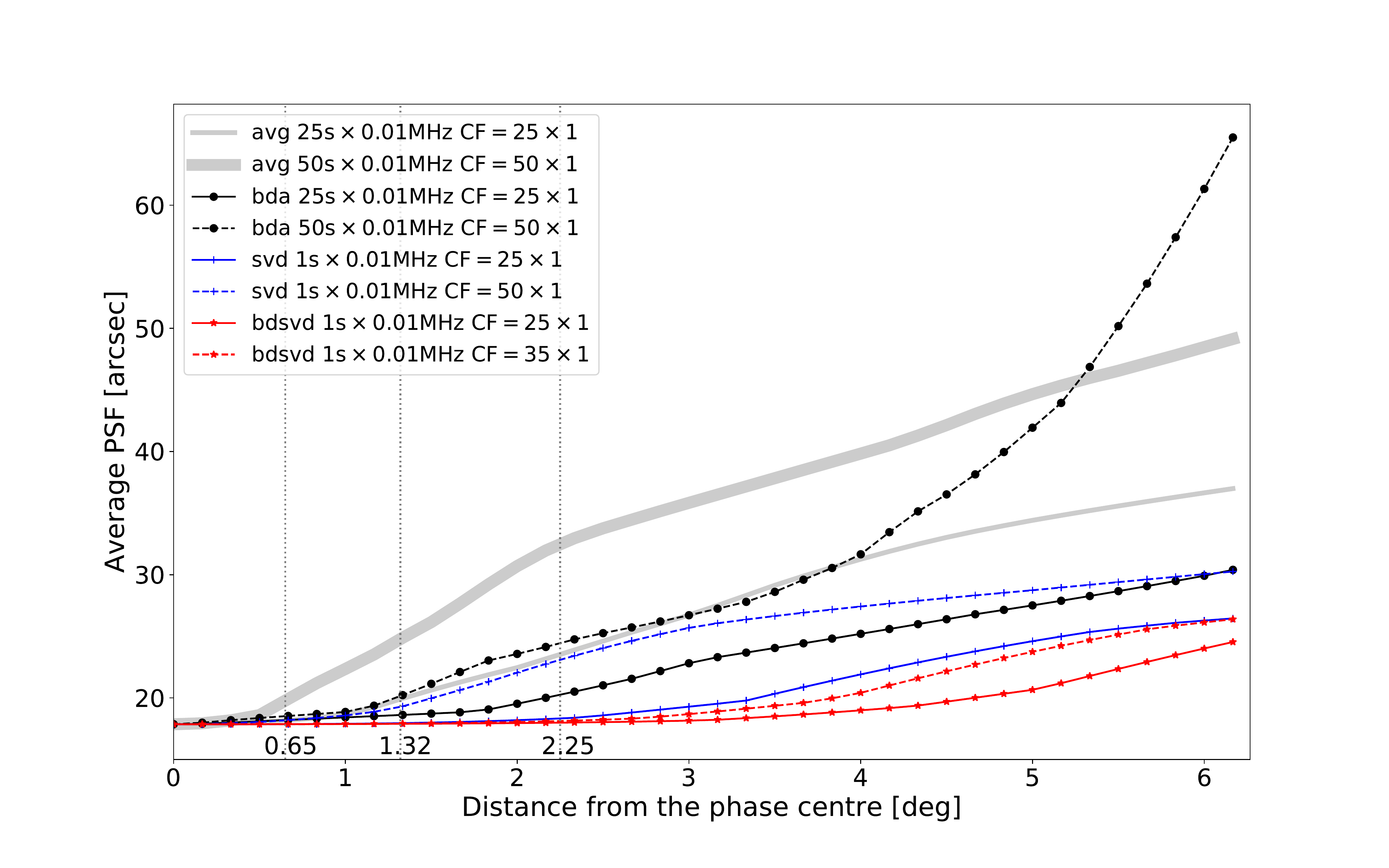}
\caption{MeerKAT telescope observing at 1.4 GHz during $166$ min $40$ s with  a total bandwidth of 0.8 MHz. The average of the radial and tangential PSF resolution measured at the FWHM are shown at various sky positions when  traditional averaging and BDA with $CF=25 \times 1$ and $CF=50 \times 1$ and for  \textit{simple SVD}, \textit{BDSVD} with $CF=25 \times 1$, $CF=35 \times 1$  and $CF=50\times 1$ are applied. The PSF is not distorted at the edges of the FoV when the SVD-related methods applied.}
\label{fig:psfdistortion2}
\end{figure*}

\begin{figure*}
\centering
  \includegraphics[width=0.92\linewidth]{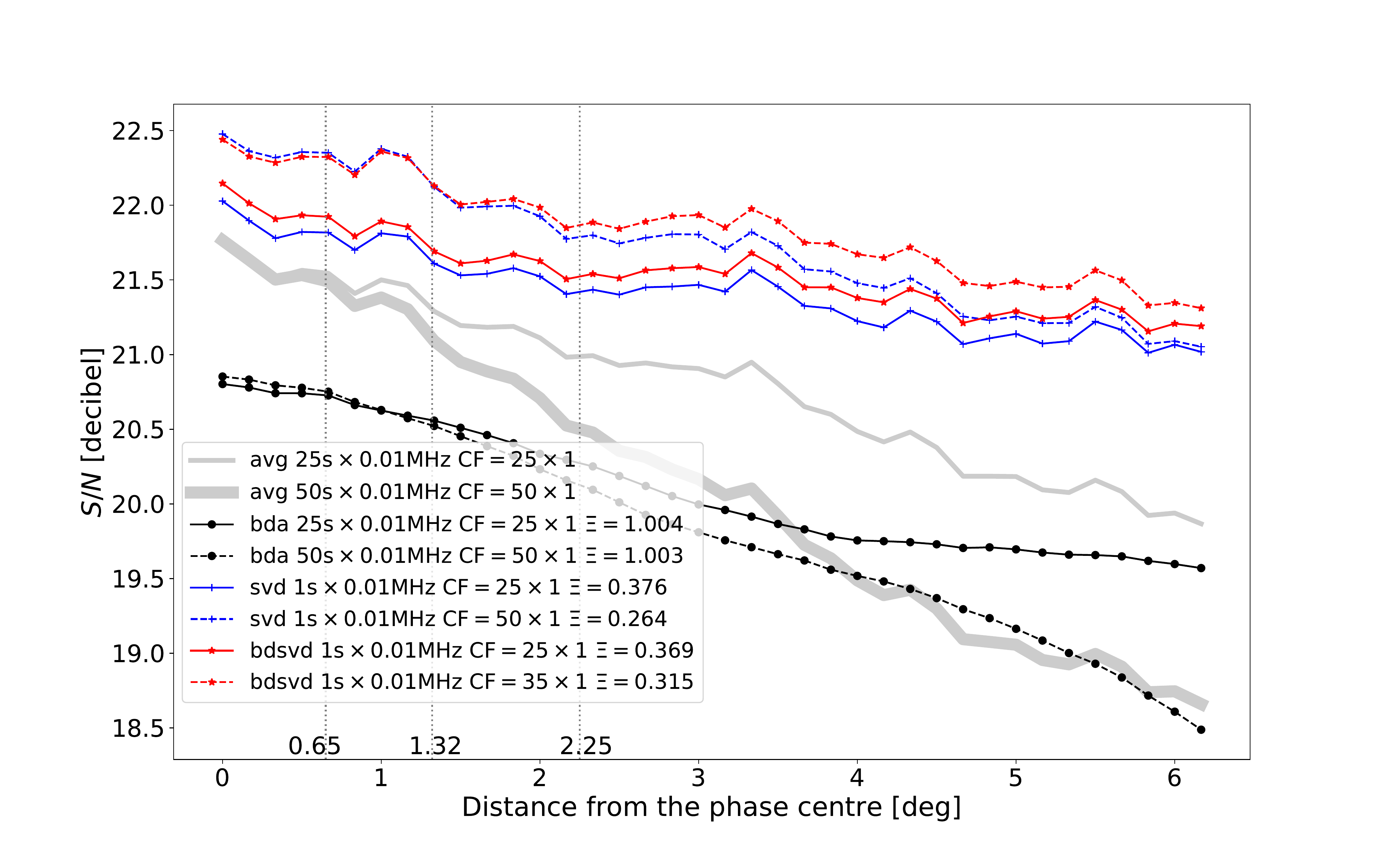}
\caption{MeerKAT telescope observing at 1.4 GHz during $166$ min $40$ s with a total bandwidth of 0.8 MHz. The S/N as defined in Eq~\ref{SNR} is measured at various sky positions for compression factors $CF=25 \times 1$, $CF=35\times 1$ and $CF=50 \times 1$ with   traditional averaging, BDA, \textit{simple SVD} and \textit{BDSVD}. The SVD-related methods offer superior S/N.}
\label{fig:snr}
\end{figure*}
As discussed in Section~\ref{Sect:introduction}, a negative aspect of averaging the visibility data is the distortion of the PSF for which the longer baselines are the major contributors. The amplitude of the PSF at a given position in the sky provides a measure of the signal loss while its width (say at the FWHM) describes how widely the source is spread out at that position in the sky. In this section, we evaluate the PSF distortion  by measuring the width of the PSF at the FWHM  when each of the discussed compression methods are applied in time. The simulated high-resolution dataset  of $1s\times 80kHz$ bins discussed in Section~\ref{MeerkATTel} is reused where the three compression factors  $CF=25\times 1$, $CF=35\times 1$ and $CF=50\times 1$ are compared for all the different methods. 

Results are shown in Figures~\ref{fig:psfdistortion1} and  \ref{fig:psfdistortion2}. Figure~\ref{fig:psfdistortion1} depicts the PSF of a source at $2.25$ deg when the high-resolution dataset is compressed for $CF=25\times 1$ with \textit{BDSVD} (top left), \textit{simple SVD} (top right), BDA (middle left)  and  traditional averaging (middle right). Figure~\ref{fig:psfdistortion1} displays the vertical (bottom left) and horizontal (bottom right) directions of the normalised PSFs as well. It is clear from this visual inspection that the width of the PSF is different for all  the compression methods, with  traditional averaging having a wider PSF compared to the other methods. Although BDA is a potential compression method that retains amplitude loss, the resulting PSF shape differs completely from that of  traditional averaging and the SVD-related methods. 

In order to quantify the width of the PSFs, Figure~\ref{fig:psfdistortion2} displays  the average of the radial and tangential PSF resolutions measured at the FWHM against distance from the phase centre of the observation.  \textit{BDSVD} shows excellent capabilities that maintain the PSF without any distortion up to a radius greater than $3$ deg with $CF=25\times 1$ and $CF=35 \times 1$. A similar result is observed for \textit{simple SVD} with $CF=25 \times 1$. However, for \textit{simple SVD} the PSF distortion starts at $\sim 1.2$ deg radius for $CF=50\times 1$. When traditional averaging or BDA is in action, the PSF quickly begins to observe distortion starting at around $\sim 1.2$ deg for each of the compression factors. These  results confirm that \textit{BDSVD} and \textit{simple SVD} can maintain the PSF without any distortion up to above a radius of $2.25$ deg; the maximum radius that the MeerKAT telescope at 1.4 GHz is capable to achieve. 

\subsubsection{Relative S/N}
The S/N as discussed theoretically in Section~\ref{noise:filtering} is shown in Figure~\ref{fig:snr} as a function of source position in the sky. The  simulation in Section~\ref{sec:sourceamp} is used to measure the source amplitude when the compression methods are applied with the different compression factors. 

Two simulations are included in this section to measure the thermal noise and noise from sources outside the FoV. The high-resolution dataset as described above is used to populate i) an empty sky with 1 Jy thermal noise and ii) a 10 Jy source is simulated at 20 deg away from the phase centre, and the different compression methods are applied separately to each  simulation. The far-field contamination is measured from  images of size $2^{10}\times 2^{10}$. Note that apart from the signal from the far-field source, this image should be empty.  \textit{BDSVD} and \textit{simple SVD} benefit from a strong improvement of  S/N to about $\sim 1.5$ dB at $2.5$ deg compared to  traditional averaging. However, BDA does not improve the S/N compared to  traditional averaging;  this is understood since thermal noise reduction scales are the same for BDA and  traditional averaging, while BDA does not suppress sources that are outside the FoV compared to  traditional averaging.
\begin{figure*}
\centering
  \includegraphics[width=0.92\linewidth]{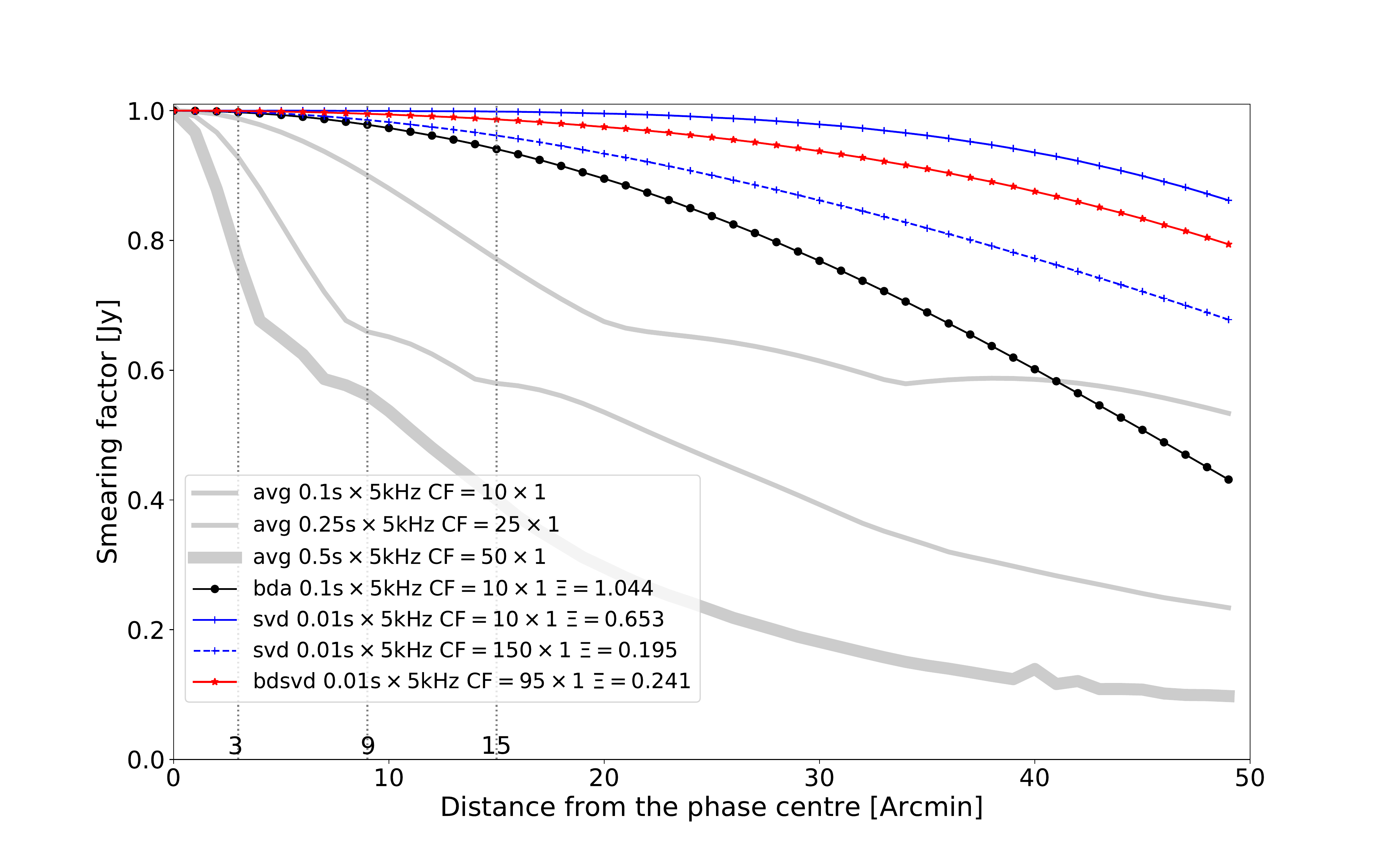}
\caption{EVN telescope observing at 1.6 GHz during $15$ min with a total bandwidth of $50$ kHz; demonstrating the degree of smearing for a 1 Jy  source amplitude loss at various sky positions  and the associated noise penalty. Smearing against source distance from the phase centre, for  traditional averaging, BDA,  \textit{simple SVD} and \textit{BDSVD} with $CF=10 \times 1$,  $CF=25 \times 1$, $CF=50\times 1$, $CF=95\times 1$ and $CF=150\times 1$. The  noise penalty $\Xi$ is given relative to the  traditional averaging bins.}
\label{fig:evntime}
\end{figure*}

\subsection{EVN}
We investigate the application of \textit{simple SVD}  and \textit{BDSVD} in VLBI to keep decorrelation down to a certain level while significantly compressing the data. We simulate a 15 min full EVN (i.e. Badary, Effelsberg, Hartebeesthoek, Jodrell
Bank, Medicina, Noto, Onsala, Shanghai, Svetloe, Torun,
Westerbork, Zelenchukskaya) at 1.6 GHz observation. With a total bandwidth of $50$ kHz channelised into 10 channels each of width $5$ kHz, the 15 min observation is sampled at each $0.01$ s which gives $90~000$ time steps. To apply the \textit{simple SVD} and \textit{BDSVD}, the $90~000$ time steps are scaled to $300\times 300$, which allows us to investigate compression factors of $CF= 10\times 1$, $CF =95\times 1$ and $CF =150\times 1$. A single point source with 1 Jy amplitude is simulated and the amplitude of the source is measured and compared to that of  traditional averaging and BDA. The results are shown in Figure~\ref{fig:evntime}.  It is observed that for a smearing factor of $1\%$ and  $CF=10\times 1$, traditional averaging and BDA would result in a FoV of 6  arcmin and 18 arcmin, respectively.  Whereas with $CF=95\times 1$, \textit{BDSVD} can image a FoV of above 30 arcmin with a superior noise reduction capability. Although these results give a good taste in the VLBI regime, the implementation of the different compression methods on real data is imperative.

\section{Conclusions}
\label{conclusion}

One of the major contributors to the large data volumes is the long baselines of an array as they influence the degree of data sampling required to avoid decorrelation of the astrophysical signal. With BDA, the sampling rate is baseline-dependent  as short baselines can be sampled far more coarsely leading to a high compression rate. BDA is an established technique for compressing radio interferometric data. However, this technique results in irregularly sampled data which, while technically supported by the MS format via variable frequency bins supported by many spectral windows, requires the data to be restructured in ways that reduce performance when processing. This work shows an approach that uses a low-rank matrix approximation to achieve greater compression rates while significantly minimising smearing compared to BDA or  traditional averaging. What is even more exciting is that when combined the low-rank approximation and  baseline-dependent formalism,  \textit{BDSVD}, effectively eliminates smearing within the FoV for the MeerKAT telescope and the EVN. However, this method has three caveats.

Firstly, SVD is computationally expensive and may be impractical for large datasets when performed on every baseline independently. Fortunately,  we show that the computations can be distributed across multiple computer nodes as discussed in Algorithms~\ref{algo:3}.

Secondly, \textit{simple SVD} and \textit{BDSVD} do not affect current calibration algorithms or software. The possibility of calibrating  visibility data remains since the compressed data is decompressed before it is calibrated.  This implies that the temporal and frequency resolutions are recovered  and consequently, do not modify the solution intervals required for calibration. We also note that the capability of current imaging software is maintained. A very important result to note is that linear transformation between the compressed data, the compressed noise and the source is maintained, thus showing that it is possible to image the sky in the compressed domain if $\I_{pqk}^\mathrm{d}$ is well understood.  This opens a potential research investigation that could possibly see the data  be imaged in the SVD space.

Thirdly, the implementation discussed in Section~\ref{sec:implementation} describes a method which compresses BDA data to an MS-compatible format on both disk and in memory.
By contrast the SVD compression schemes described here produce singular values that must be expanded "on-the-fly" to full resolution in memory.
There is, therefore, an interesting contrast between the two approaches as the first decreases both the amount of data and computation (FLOPs) performed by algorithms on that data.
This is achieved at the cost of implementation complexity, especially  with regards to  calibration techniques which must reason about solution intervals that lie across
multiple data points on the sparse domain in which BDA data lies. \textit{Simple SVD} and \textit{BDSVD} offer superior compression to BDA and  traditional averaging, but requires processing data at full data resolution, for reduced implementation complexity.
Decompressing data "on-the-fly" is already offered in the PyDATA ecosystem by packages such as BLOSC \citep{haenel14blosc}, which compresses data into chunks and decompresses them exactly into the L1 cache of a CPU.
This reduces the memory intensity of an algorithm, which is beneficial since modern CPUs are starved for data by their memory system \citep{alted10cpustarve} and radio interferometric data is very large.
In such a regime, it may be possible to transmit the singular values into the L1 cache of
a CPU and expand the data to full resolution for consumption by algorithms. Additional possible future work would investigate an online compression-decompression strategy  using the full archiving potential of SVD-related  methods.

Real data that is contaminated with Radio Frequency Interference (RFI),  power is distributed across all visibilities when decompressing the SVD. There are two ways to avoid this situation: either the RFIs are removed from the data before applying the SVD or the SVD is used to remove the RFI during compression. The latter requires intense investigation as future work, since removing the RFI is equivalent to removing the higher singular value components. Removing higher singular value components is problematic because it is unclear where weak radio signals will begin to be suppressed in the process.

Real data usually has a few entries flagged for some reason. Interesting future work would be to investigate how these flagged entries  affect singular values. We expect that the situation will fall somewhere between the two below scenarios:
\begin{itemize}
    \item If the flagged real  data entries belong to the same neighbourhood and are assigned the same value; this increases the similarity in the neighbourhood and hence SVD will result in a lower rank  (consequently, the compression factor will be bigger) compared to no flagging; and
    \item If the flagged real data entries are not from the same neighbourhood (e.g. randomly flagged data), the rank of the SVD will be higher (consequently, the compression factor will be smaller) compared to no flagging.
\end{itemize}
To address this problem, a potential solution would be to perform in-painting on the model data where flags are present and then use this in-painted data to fill in the flagged entries in the uncompressed data prior to applying the SVD.

\section*{Acknowledgements}
This work is based upon research supported by the South African Research Chairs Initiative of the Department of Science and Technology and National Research Foundation.
The MeerKAT telescope is operated by the South African Radio Astronomy Observatory, which is a facility of the National Research Foundation, an agency of the Department
of Science and Innovation. MA acknowledges support from Rhodes University.

\section*{Data Availability}
The data underlying this article will be shared on reasonable
request to the corresponding author



\bibliographystyle{mnras}
\bibliography{example} 




\appendix

\begin{appendix} 
\section{Keeping track of the noise}
\label{ap:A}%
Given an observation (assuming $M<N$)
\begin{equation}
    \underset{M\times N}{Y} = \bar{Y} + \epsilon = \underset{M\times M}{U} \underset{M\times M}{\Lambda} \underset{M\times N}{V},
\end{equation}
with
\begin{equation}
    \mbox{vec}(\epsilon) \quad  \sim \quad N(0, \Sigma).
\end{equation}
\textit{Can we keep track of the noise properties after thresholding the singular values?} We can use the decomposition to keep track of the covariance of the singular values. To do this we note that, since $U$ and $V$ are unitary matrices, we have
\begin{equation}
    \Lambda = U^\dagger(\bar{Y} + \epsilon)V^\dagger = \bar{\Lambda} + \epsilon_\Lambda,
\end{equation}
with $\bar{\Lambda} = U^\dagger \bar{Y} V^\dagger$ and $\epsilon_\Lambda = U^\dagger \epsilon V^\dagger$. Now the covariance in SVD space is given by 
\begin{eqnarray}
    \Sigma_\Lambda &=& \mathbb{E}[\mbox{vec}(\epsilon_\Lambda) \mbox{vec}(\epsilon_\Lambda^\dagger)] \\
    &=& \mathbb{E}[(V \otimes U^\dagger) \mbox{vec}(\epsilon) \mbox{vec}(\epsilon)^\dagger (V \otimes U^\dagger)^\dagger] \\
    &=& (V \otimes U^\dagger) \mathbb{E}[\mbox{vec}(\epsilon) \mbox{vec}(\epsilon)^\dagger] (V \otimes U^\dagger)^\dagger \\
    &=& (V \otimes U^\dagger) \Sigma (V \otimes U^\dagger)^\dagger,
\end{eqnarray}
where we have used the identity $\mbox{vec}(ABC) = (C^\dagger \otimes A) \mbox{vec}(B)$ and the fact that
\begin{equation}
    \mathbb{E}[\mbox{vec}(\epsilon) \mbox{vec}(\epsilon)^\dagger]  = \Sigma. 
\end{equation}
Notice that, from the mixed product property of the Kronecker product, the noise in SVD space will be i.i.d. whenever $\epsilon$ is i.i.d (i.e. $\Sigma$ is a multiple of the identity). Keeping track of the uncertainty after thresholding is trivial when this is the case because the singular values are independent. Things are a bit more complicated when there are off diagonal entries in $\Sigma_\Lambda$. It is likely that the $\Sigma_\Lambda$ will be diagonally dominant when $\Sigma$ is diagonal but this has to be checked. If it is we can probably safely just treat it as a diagonal matrix.
\end{appendix}


\bsp	
\label{lastpage}
\end{document}